\documentclass[10pt,A4paper]{article}
\usepackage[T1]{fontenc}
\usepackage[utf8]{inputenc}
\usepackage{authblk}
\usepackage{arxiv}

\usepackage{amsmath,amssymb,amsfonts}
\usepackage{url}

\newcommand{\bs}[1]{\boldsymbol{#1}}

\newcommand{\dpar}[2]{\frac{\partial #1}{\partial #2}}
\newcommand{\diff}[2]{\frac{d #1}{d #2}}
\newcommand{\ddiff}[2]{\frac{\mathsf{D} #1}{\mathsf{D} #2}}

\newcommand{\mysecref}[1]{Section~\ref{#1}}
\newcommand{\myeqref}[1]{Eq.~(\ref{#1})}
\newcommand{\myfigref}[1]{Fig.~\ref{#1}}
\newcommand{\myalgref}[1]{Algorithm~\ref{#1}}
\newcommand{\mytabref}[1]{Table~\ref{#1}}

\usepackage{tikz}
\usetikzlibrary{positioning}
\usetikzlibrary{decorations.pathmorphing,patterns}
\usetikzlibrary{calc,patterns,decorations.markings}
\usetikzlibrary{shapes,arrows}

\usepackage{pgfplots}
\usepackage{algorithm,algpseudocode}
\usepackage{algorithmicx}
\usepackage{amsmath}
\usepackage{amsfonts}
\usepackage{gensymb}
\usepackage{siunitx}
\usepackage[colorlinks,bookmarksnumbered,bookmarksopen,citecolor=blue,urlcolor=blue,linkcolor=blue,]{hyperref}

\usepackage{xcolor}

\title{Deep learning of thermodynamics-aware reduced-order models from data}

\author[1]{Quercus Hern\'andez}
\author[1]{Alberto Bad\'ias}
\author[1]{David Gonz\'alez}
\author[2]{Francisco Chinesta}
\author[1]{El\'ias Cueto}
\affil[1]{{\small Aragon Institute of Engineering Research. Universidad de Zaragoza. Zaragoza, Spain.}}
\affil[2]{{\small ESI Group chair. PIMM Lab. ENSAM Institute of Technology. Paris, France.}}

\begin{document}

\maketitle

\begin{abstract}
We present an algorithm to learn the relevant latent variables of a large-scale discretized physical system and predict its time evolution using thermodynamically-consistent deep neural networks. Our method relies on sparse autoencoders, which reduce the dimensionality of the full order model to a set of sparse latent variables with no prior knowledge of the coded space dimensionality. Then, a second neural network is trained to learn the metriplectic structure of those reduced physical variables and predict its time evolution with a so-called structure-preserving neural network. This data-based integrator is guaranteed to conserve the total energy of the system and the entropy inequality, and can be applied to both conservative and dissipative systems. The integrated paths can then be decoded to the original full-dimensional manifold and be compared to the ground truth solution. This method is tested with two  examples applied to fluid and solid mechanics.
\end{abstract}

\section{Introduction}

Physical simulation has become an indispensable tool for engineers to recreate the operative conditions of a mechanical system and make decisions about its optimal design, ranging from composite building structures to complex fluid-solid interaction CFD simulations. These phenomena are often discretized in fine meshes resulting in millions of degrees of freedom, which are computationally expensive to handle, but their solutions are contained in lower-dimensional spaces. This is the so-called manifold hypothesis \cite{manifold}.

Thus, several methods try to overcome this inconvenience by reducing the dimensionality of the problem, computing a suitable reduced basis and projecting the full order model on it. The very first projection-based model order reduction (MOR) methods relied on linear transformations with some additional constraints, such as Proper Orthogonal Decomposition (POD) \cite{niroomandi2008real,du2013pod}, Reduced-Basis technique \cite{prud2002reliable} or Galerkin projection \cite{rowley2004model,farrell2011conservative}. However, these linear mappings are only locally accurate, so they fail in modeling more complex nonlinear phenomena and sometimes require prior information about the governing equations of the problem physics.

In order to overcome these limitations, several techniques have been developed in the machine learning framework that provide nonlinear mappings, such as Locally Linear Embedding  \cite{badias2019augmented}, Topological Data Analysis \cite{moya2019learning}, kernel Principal Component Analysis \cite{moya2020physically} or Neural Networks, by means of Autoencoders \cite{goodfellow2016deep}. In the present work we focus on this last method, which has proven to learn highly nonlinear manifolds in a wide variety of fields such as physics \cite{farina2020searching}, chemistry \cite{liu2018constrained}, mechanics \cite{lee2020model} or computational imaging \cite{marco2017deeptof}. Autoencoders used as a model reduction tool, project the original data (assumed to form a high-order manifold) to a reduced manifold. However, most of the current works rely on prior knowledge, or parametric search, of the optimal latent dimensionality of the problem. Here lies one of the key concepts of our method, which is able to learn a sparse representation of the latent space within a given reconstruction error bound.

These same machine learning tools can be used to learn the underlying physics of the problem. Very often, neural networks have been criticized for constituting a sort of black box, whose results---besides needing a big amount of data---are unpredictable. Therefore, adding previous knowledge on the physics of the problem helps to ensure the physical meaning of the results, while keeping to a minimum the amount of data needed for successful predictions. Several authors have developed frameworks for solving nonlinear PDEs with accurate results \cite{raissi2019physics,kelly2020easynode}. However, they require information about the system nature and governing equations, which are usually unknown. Some methods bypass this problem by learning energetic invariants of the system \cite{bertalan2019learning,greydanus2019hamiltonian,toth2019hamiltonian} or exploiting the symplectic structure of the problem \cite{zhong2019symplectic,tong2020symplectic,jin2020learning}, reporting promising and interpretable results for Hamiltonian dynamics. Nonetheless, few methods are valid for dissipative effects such as friction, heat dissipation or plasticity, which are usually found in real life engineering problems.

The authors already presented a methodology to learn the time evolution of general physical systems by enforcing the GENERIC (an acronym of General Equation for the Non-Equilibrium Reversible-Irreversible Coupling) structure of the problem \cite{ottinger1997dynamics, grmela1997dynamics}, with the so-called Structure-Preserving Neural Networks \cite{hernandez2020structure}. This networks result in a thermodynamically-consistent integrator that is valid for both conservative (Hamiltonian) and dissipative systems. However, these networks operate only on full-order descriptions of the system, resulting in a costly procedure with limited engineering applicability for systems of tens of thousands to millions of degrees of freedom. The aim of this work is to apply this algorithm to more complex dynamical systems, combined with the nonlinear model order reduction power of autoencoders. The proposed methodology is a completely general method that is able to unveil the true effective dimensionality of the sampled data with no user intervention, and to construct from it a reduced-order integrator of the dynamics of the system with no previous knowledge on the nature of the system at hand. The resulting full-order reconstructions of the dynamics are guaranteed to conserve energy and dissipate entropy, as dictated by the laws of thermodynamics.

The outline of the paper is as follows. A brief description of the problem setup is presented in \mysecref{sec:statement}. Next, in \mysecref{sec:method}, the methodology is presented of both the autoencoder model order reduction and the GENERIC formalism used to solve the stated problem. Two validation examples are reported: a Couette flow in a viscolastic fluid (\mysecref{sec:visco}) and a rolling hyperelastic tire (\mysecref{sec:tire}). The paper is then completed with a discussion in \mysecref{sec:conc}.

\section{Problem Statement}\label{sec:statement}

In this work we exploit the so-called ``dynamical systems equivalence'' of machine learning \cite{E2017}. Consider a system whose governing variables will be hereafter denoted by $\bs z \in \mathcal M \subseteq \mathbb R^D$, with $\mathcal M$ the state space of these variables, which is assumed to have the structure of a differentiable manifold in $\mathbb R^D$. The full-order model of a  given physical phenomenon can be expressed as a system of differential equations encoding the time evolution of a set of governing variables $\bs z$,
{
\begin{equation}\label{eq:pde}
\dot{\bs z}= \diff{\bs{z}}{t} = F(\bs{z},t),\; t\in\mathcal{I}=(0,T],\; \bs z(0)=\bs z_0,
\end{equation}}
where {{$\bs{q}$ and $t$ refer to the space and time coordinates within a domain with $n=2,3$ dimensions} $t$ refers to the time coordinate}. The objective of the learning procedure is, therefore, to find {$F(\bs z, t)$}, the function that gives, after a prescribed time horizon $T$, the flow map $\bs z_0 \rightarrow \bs z(\bs z_0,T)$.

The dimensionality reduction technique, in addition, seeks a simplified representation of the full-order state vector $\bs{z}$ through a set of latent, reduced variables $\bs x \in \mathcal N \subseteq \mathbb R^d$ contained in a trial manifold with reduced dimensionality, lower than the original space $\mathcal M$. The mapping between both spaces can be denoted by $\phi : \mathcal M \subseteq \mathbb R^D \rightarrow \mathbb R^d$ with $d \ll D$. Similarly, the inverse mapping $\phi^{-1}$ allows to undo the transformation, returning to the original full-order space.

The goal of this paper is to find the convenient mapping $\phi$ for a dynamical system governed by \myeqref{eq:pde} in order to efficiently learn the underlying physics in the reduced space $\mathcal N$ and then predict its time evolution. The solution is forced to fulfil the basic thermodynamic requirements of energy conservation and entropy inequality restrictions via the GENERIC formalism.

\section{Methodology}\label{sec:method}

\begin{figure}[h]
   \centering      
   \includegraphics[width=\textwidth]{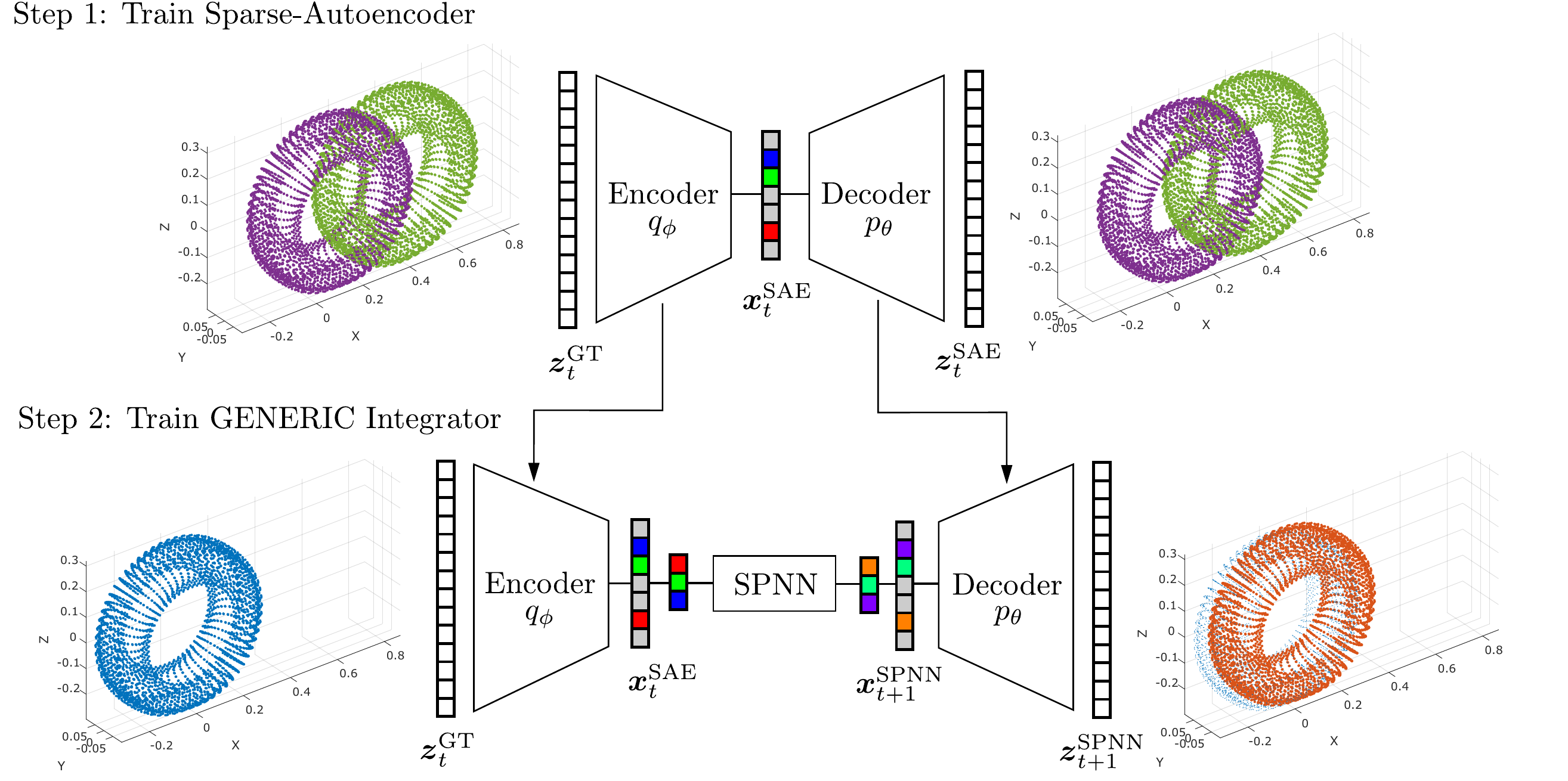}      
 \caption{Block diagram of the proposed algorithm. Snapshots of the rolling tire problem, see \mysecref{sec:tire}, have been included for illustration purposes. Step 1: A sparse autoencoder (SAE) is trained with time snapshots of a ground truth physical simulation, in order to learn an encoded representation of the full-order space. Step 2: A structure-preserving neural network (SPNN) is trained to integrate the full time evolution of the latent variables, consistently with the GENERIC structure of the underlying physics of the problem.}
 \label{fig:method}
\end{figure}

The proposed algorithm divides the problem in two main steps, sketched in \myfigref{fig:method}. First, the full order model is encoded to a reduced manifold with a nonlinear mapping via an autoencoder \cite{lee2020model}. This autoencoder learns a latent representation of a state vector of a physical system, in order to handle a wide amount of simulation data in a compact form. The full order simulation data presented in this work is generated in silico, but the same procedure could be applied to measured data in a real physical system. 

Secondly, a structure-preserving neural network \cite{hernandez2020structure} is trained with several snapshots of the physical simulation. { This net works as an integrator which predicts the time evolution of the system within the GENERIC formalism \cite{grmela1997dynamics,ottinger1997dynamics}.} This integration scheme preserves the thermodynamic structure of the latent variables in the reduced manifold \cite{ottinger2015preservation} ensuring, as we said, the basic laws of thermodynamics of energy conservation and entropy inequality. These integrated variables are then projected back to the original manifold of the full order model with the decoder.

\subsection{Model Reduction with Sparse-Autoencoders}

An autoencoder is a type of artificial neural network which reduces the dimensionality of an input into a coded version, which ideally contains the same information, by learning the identity function. It is composed by an encoder $q_\phi$, which maps high-dimensional data $\bs{z}\in\mathbb{R}^D$ onto a low-dimensional code $\bs{x}\in\mathbb{R}^d$ with $d\ll D$, and a decoder $p_\theta$, which applies the inverse mapping back to the original full-order manifold,
\begin{eqnarray}\label{eq:encoder}
q_\phi:\mathbb{R}^D\rightarrow \mathbb{R}^d,\quad \bs{x}=q_\phi(\bs{z}),\\\label{eq:decoder}
p_\theta:\mathbb{R}^d\rightarrow \mathbb{R}^D,\quad \hat{\bs{z}}=p_\theta(\bs{x}).
\end{eqnarray}

The vector $\bs{z}$ is often referred as the full order vector, whereas its coded vector $\bs{x}$ is referred as \textit{code} or \textit{latent variable}. In this work, we use a bottleneck architecture composed by several stacked fully-connected hidden layers $N_h$ in both the encoder and decoder. Each layer is modelled as a multilayer perceptron (MLP), which is mathematically defined as
\begin{equation}\label{eq:forward_net}
 \bs{x}^{[l]} = \sigma(\bs{w}^{[l]}\bs{x}^{[l-1]}+\bs{b}^{[l]}),
\end{equation}
where $l$ is the index of the current layer, $\bs{x}^{[l-1]}$ and $\bs{x}^{[l]}$ are the layer input and output vector respectively, $\bs{w}^{[l]}$ is the weight matrix, $\bs{b}^{[l]}$ is the bias vector and $\sigma$ is the activation function. The activation functions are usually nonlinear, allowing the encoding and decoding of complex nonlinear phenomena by stacking several layers together.

The latent vector dimensionality $d$ in \myeqref{eq:encoder} is, a priori, unknown. Thus, we add a sparsity condition to the bottleneck to force the autoencoder to learn the number of latent variables needed to encode the necessary information of the full order model. Even if the latent layer has a fixed number of units $N_d$, the sparsity penalizer is able to find (at least a good approximation to) the intrinsic dimensionality of the low-dimensional data $\bs{x}$. Here, no prior on the reduced dimension is needed. Thus, further in this text, the autoencoder with sparsity regularization is referred as sparse autoencoder (SAE).

The loss function for our neural network is composed of two different terms:
\begin{itemize}
\item \textbf{Reconstruction loss}: This term minimizes the difference between the ground truth vector $\bs{z}_n^{\text{GT}}$ and the autoencoder reconstruction $\bs{z}_n^{\text{SAE}}$ in the snapshot $n$. { This enforces the network to learn the identity function,
\begin{equation}\label{eq:SAE_rec}
\mathcal{L}_n^{\text{rec}}=||\bs{z}_n^{\text{GT}}-\bs{z}_n^{\text{SAE}}||_2^2.
\end{equation}}

\item \textbf{Regularization}: In order to impose the sparsity of the latent vector, several regularizers can be used \cite{ng2011sparse}. Due to the continuous nature of the physical data, it is found more convenient to use L1-norm penalizer, which enforces hard zeros in the latent variables that are not relevant, 
\begin{equation}\label{eq:SAE_reg}
\mathcal{L}_n^{\text{reg}}=\sum_{i=1}^{N_d}|\bs{x}_i^{\text{SAE}}|.
\end{equation}

\end{itemize}

The temporal snapshots of the physical simulations are split in a partition of train snapshots ($N_{\text{train}}=80\%$ of the database snapshots) and test snapshots ($N_{\text{test}}=20\%$ of the database snapshots) so that $N_T=N_{\text{train}}+N_{\text{test}}$. The total loss function is computed as the mean squared error (MSE) of the data reconstruction loss and the sparsity regularization term for the train snapshots ($N_{\text{train}}$). The sparsity loss is multiplied by a regularization hyperparameter $\lambda_r^{\text{SAE}}$, which is responsible for the trade-off between the reconstruction fidelity of the autoencoder and the sparsity of the latent vector $\bs{x}$,
\begin{equation}\label{eq:SAE_loss}
\mathcal{L}^{\text{SAE}}=\frac{1}{N_{\text{train}}}\sum_{n=0}^{N_{\text{train}}}(\mathcal{L}_n^{\text{rec}}+\lambda_r^{\text{SAE}}\mathcal{L}_n^{\text{reg}}).
\end{equation}

The backpropagation algorithm \cite{paszke2017automatic} is then used to calculate the gradient of the loss function for each encoder and decoder parameters $\phi$ and $\theta$ (weight and bias vectors of both blocks), which are updated with the gradient descent technique \cite{ruder2016overview}. An overview of the training algorithm of the SAE is sketched in \myalgref{alg:SAE_train}.

\begin{algorithm*}[h!]
\caption{Pseudocode for the training algorithm of the Sparse-Autoencoder.}\label{alg:SAE_train}
\begin{algorithmic}
  \State \textbf{Load database:} $\bs{z}^{\text{GT}}$ (train partition);
  \State \textbf{Define network architecture:} $N_{\text{in}}^{\text{SAE}}=N_{\text{out}}^{\text{SAE}}=D$, $N_{h}^{\text{SAE}}$, $N_{d}^{\text{SAE}}$, $\sigma_j^{\text{SAE}}$;
  \State \textbf{Define hyperparameters:} $l_r^{\text{SAE}}$, $\lambda_r^{\text{SAE}}$;
  \State Initialize $w_{i,j}^{\text{SAE}}$, $b_j^{\text{SAE}}$;
  \For{each epoch}
  		\State Initialize loss function: $C=0$;
  		\For{each train snapshot}
  			\State Encoder: $\bs{x}_n^{\text{SAE}}=q_{\phi}(\bs{z}^{\text{GT}}_n)$; \Comment \myeqref{eq:encoder}
  			\State Decoder: $\bs{z}_n^{\text{SAE}}=p_{\theta}(\bs{x}^{\text{SAE}}_n)$; \Comment \myeqref{eq:decoder}
  			\State Loss function: $C \gets C + \mathcal{L}_n^{\text{rec}} + \lambda_r^{\text{SAE}}\mathcal{L}_n^{\text{reg}}$; \Comment \myeqref{eq:SAE_rec}, \myeqref{eq:SAE_reg}
  		\EndFor		
  		\State MSE loss function: $\mathcal{L}^{\text{SAE}} \gets \frac{C}{N_{\text{train}}}$ \Comment \myeqref{eq:SAE_loss}
  		\State Backward propagation;
  		\State Optimizer step;
  \EndFor
\end{algorithmic}
\end{algorithm*}

The SAE performance is then evaluated with the mean squared error (MSE) of the test snapshots ($N_{\text{test}}$) for each state variable ($\bs{z}$),
\begin{equation}\label{eq:SAE_error}
\text{MSE}^{\text{SAE}}\;(\bs{z})=\frac{1}{N_{\text{test}}}\sum_{n=0}^{N_{\text{test}}}\varepsilon_n^2=\frac{1}{N_{\text{test}}}\sum_{n=0}^{N_{\text{test}}} \left(\bs{z}^{\text{GT}}_{n}-\bs{z}^{\text{SAE}}_{n}\right)^2,
\end{equation}
tested with two different databases of nonlinear systems. A pseudocode of the testing process of the SAE is shown in \myalgref{alg:SAE_test}.

\begin{algorithm*}[h!]
\caption{Pseudocode for the test algorithm of the Sparse-Autoencoder.}\label{alg:SAE_test}
\begin{algorithmic}
  \State \textbf{Load database:} $\bs{z}^{\text{GT}}$ (test partition);
  \State \textbf{Load network parameters};
  		\For{each test snapshot}
  			\State Encoder: $\bs{x}_n^{\text{SAE}}=q_{\phi}(\bs{z}^{\text{GT}}_n)$; \Comment \myeqref{eq:encoder}
  			\State Decoder: $\bs{z}_n^{\text{SAE}}=p_{\theta}(\bs{x}^{\text{SAE}}_n)$; \Comment \myeqref{eq:decoder}
  			\State Compute Squared Error: $\varepsilon_n^2=\left(\bs{z}_n^{\text{GT}}-\bs{z}_n^{\text{SAE}}\right)^2$; \Comment \myeqref{eq:SAE_error}
  		\EndFor
  \State Compute $\text{MSE}^{\text{SAE}}\;(\bs{z})$; \Comment \myeqref{eq:SAE_error}	
\end{algorithmic}
\end{algorithm*}

Once the problem is reduced to a lower-dimensional manifold, a second neural network can be trained to learn the underlying physics of the problem, being able to integrate the whole simulation trajectory with thermodynamic consistency. This is achieved by using a structure-preserving neural network \cite{hernandez2020structure}, and is explained in the next section.

\subsection{The GENERIC formalism}

There are different forms of enforcing physical meaning to the results of a particular neural network. One could be the enforcement of the structure of a particular partial differential equation, as in \cite{raissi2019physics}. This is known as adding an {\em inductive bias} \cite{battaglia2018relational}. An inductive bias is a way to enforce an algorithm to prioritize one solution to another. In our case, we try to guarantee as much as possible the physical meaning of the solution, but without enforcing any particular physical law, which may be even unknown. We do this by adding a regularization term to our neural network. This regularization will enforce the fulfillment of the first and second laws of thermodynamics.

A Structure-Preserving Neural Network \cite{hernandez2020structure} (from now on, SPNN) is a type of artificial neural network that learns the metriplectic structure of a general dynamical system \cite{morrison1986paradigm}, with both conservative and dissipative phenomena, by imposing a GENERIC structure \cite{ottinger1997dynamics, grmela1997dynamics}. 

In this approach, the reversible or conservative contribution is assumed to be of Hamiltonian form, requiring an energy function $E(\bs{x})$ and a Poisson bracket $\lbrace\bs{x},E\rbrace$ acting on an arbitrary state vector $\bs{x}$. Similarly, the remaining irreversible contribution to the energetic balance of the system is generated by the nonequilibrium entropy $S(\bs{x})$ with an irreversible or friction bracket $[\bs{x},S]$.

The GENERIC formulation of time evolution for nonequilibrium systems, described by a set of $\bs{x}$ state variables required for its complete description, is given by
\begin{equation}\label{eq:brackets}
\frac{d\bs{x}}{dt}=\lbrace\bs{x},E\rbrace+[\bs{x},S].
\end{equation}

For practical use, it is convenient to reformulate the brackets in two algebraic or differential operators 
$$
\bs L : T^*\mathcal M \rightarrow T\mathcal M,\quad \bs M : T^*\mathcal M \rightarrow T\mathcal M,
$$
where $T^*\mathcal M$ and $T\mathcal M$ represent, respectively, the cotangent and tangent bundles of $\mathcal M$. These operators inherit the mathematical properties of the original bracket formulation. The operator $\bs{L}(\bs{x})$ represents the Poisson bracket and is required to be skew-symmetric (a cosympletic matrix). Similarly, the friction matrix $\bs{M}(\bs{x})$ accounts for the irreversible part of the system and is symmetric and positive semi-definite. Then, the brackets of \myeqref{eq:brackets} can be replaced by their homologous matrix operators
$$
\lbrace \bs{A},\bs{B}\rbrace=\dpar{\bs{A}}{\bs{x}}\bs{L}\dpar{\bs{B}}{\bs{x}}, \qquad [\bs{A},\bs{B}]=\dpar{\bs{A}}{\bs{x}}\bs{M}\dpar{\bs{B}}{\bs{x}},
$$
resulting in the time-evolution equation for the state variables $\bs{x}$,
\begin{equation}\label{eq:generic}
 \diff{\bs{x}}{t} = \bs{L} \dpar{E}{\bs{x}} + \bs{M} \dpar{S}{\bs{x}}.
\end{equation}

This equation is completed with two degeneracy conditions $$\lbrace S,\bs{x}\rbrace=\bs{0},\quad [E,\bs{x}]=\bs{0}.$$ The first one states that the entropy is a degenerate functional of the Poisson bracket, showing the reversible nature of the Hamiltonian contribution to the dynamics. The second expression states the conservation of the total energy of the system with a degenerate condition of the energy with respect to the friction bracket. These restrictions can be reformulated in a matrix form in terms of the $\bs{L}$ and $\bs{M}$ operators, resulting in the following degeneracy restrictions:
\begin{equation}\label{eq:degen}
 \bs{L} \dpar{S}{\bs{x}}=\bs{M} \dpar{E}{\bs{x}} = 0.
\end{equation}

The degeneracy conditions, in addition to the non-negativeness of the irreversible bracket, guarantees the first (energy conservation) and second (entropy inequality) laws of thermodynamics,
\begin{equation}\label{eq:Econs}
\frac{dE}{dt}=\lbrace E,E\rbrace =0,\quad\frac{dS}{dt}=[S,S]\geq 0.
\end{equation}

\subsection{Structure-Preserving Neural Networks}

Based on this theoretical formalism, a structure-preserving neural network imposes the GENERIC thermodynamically-sound structure in discretized approach,
\begin{equation}\label{eq:generic_disc}
 \frac{\bs{x}_{n+1}-\bs{x}_{n}}{\Delta t} = \mathsf{L}_n\cdot\ddiff{E_n}{\bs{x}_n} + \mathsf{M}_n\cdot\ddiff{S_n}{\bs{x}_n},
\end{equation}
where the time derivative is substituted by a forward Euler scheme in time increments $\Delta t$, where $\bs{x}_{n+1}=\bs{x}_{t+\Delta t}$. $\mathsf{L}_n$ and $\mathsf{M}_n$ are the discretized version of the Poisson and friction operators. $\ddiff{E_n}{\bs{x}_n}$ and $\ddiff{S_n}{\bs{x}_n}$ represent the discrete gradients of the energy and the entropy.

Manipulating algebraically \myeqref{eq:generic_disc} and including the degeneracy conditions of \myeqref{eq:degen}, the proposed integration scheme for predicting the dynamics of a physical system is the following
\begin{equation}\label{eq:integration}
 \bs{x}_{n+1}=\bs{x}_{n} + \Delta t \left( \mathsf{L}_n\cdot\ddiff{E_n}{\bs{x}_n} + \mathsf{M}_n\cdot\ddiff{S_n}{\bs{x}_n} \right)
\end{equation}
subject to:
\begin{equation}\label{eq:degendisc}
 \mathsf{L}_n\cdot\ddiff{S_n}{\bs{x}_n} = 0, \quad
 \mathsf{M}_n\cdot\ddiff{E_n}{\bs{x}_n} = 0,
\end{equation}
ensuring the thermodynamical consistency of the resulting model. From now on, the energy and entropy gradients will be shortened as $\ddiff{E_n}{\bs{x}_n}\equiv\mathsf{DE}_n$ and $\ddiff{S_n}{\bs{x}_n}\equiv\mathsf{DS}_n$.

Unlike previous work \cite{hernandez2020structure}, the GENERIC structure is imposed to the reduced order model learnt by the sparse autoencoder, so there is no prior information about the $\mathsf{L}$ and $\mathsf{M}$ matrices. Instead, the SPNN is forced to automatically learn them on each learning set time step, $\mathsf{L}_n$ and $\mathsf{M}_n$, with their respective skew-symmetric and symmetric conditions. Similarly, the energy and entropy gradient, $\mathsf{DE}_n$ and $\mathsf{DS}_n$, are computed on each time step and no finite-difference approach is needed.

The structure-preserving neural network uses a feed-forward scheme \cite{schmidhuber2015deep}, consisting of several fully-connected layers with no cyclic connections. The input of the neural net is the encoded vector state of a given time step $\bs{x}_n^{\text{SAE}}=q_{\phi}(\bs{x}_n^{\text{GT}})$, and the outputs are the concatenated GENERIC matrices ($\mathsf{L}_n$, $\mathsf{M}_n$) and energy and entropy gradient matrices ($\mathsf{DE}_n$, $\mathsf{DS}_n$). Then, using the GENERIC forward integration scheme in \myeqref{eq:generic_disc}, the reduced state vector at the next time step $\bs{x}_{n+1}^{\text{SPNN}}$ is obtained.

Following \myeqref{eq:integration}, the input dimension of the SPNN is the same as the dimension of the sparsified latent variables $\bs{x}_n^{\text{SAE}}$ ($N_{\text{in}}^{\text{SPNN}}=d$). Consequently, the GENERIC matrices $\mathsf{L}_n$ and $\mathsf{M}_n$ are squared with dimension $d^2$ each, which can be reduced to {$d\cdot(d-1)/2$ and} $d\cdot(d+1)/2$ taking into account the skew-symmetric and symmetric elements respectively. { Additionally, the matrix $\mathsf{M}_n$ is assembled by taking the absolute value of the diagonal elements of the resulting lower triangular matrix and multiplying it by its transpose. By the Cholesky factorization, this ensures that $\mathsf{M}_n$ is positive semidefinite.} The energy and entropy gradient matrices $\mathsf{DE}_n$ and $\mathsf{DS}_n$ have the same dimension $d$ as the state vector. The final output dimension of the integrator network is then $N_\text{out}^{\text{SPNN}}=d\cdot(d+1)/2+d\cdot(d-1)/2+2\cdot d=d\cdot(d+2)$.

The loss function for the SPNN is composed of three different terms:
\begin{itemize}

\item \textbf{Data loss}: The main loss condition is the agreement between the network output and the real data. It is computed as the squared error sum, computed between the predicted state vector $\bs{x}_{n+1}^{\text{SPNN}}$ and the ground truth solution based on the SAE output $\bs{x}_{n+1}^{\text{SAE}}$ for each time step,
\begin{equation}\label{eq:loss_data}
\mathcal{L}_n^{\text{data}} = ||\bs{x}_{n+1}^{\text{SAE}}-\bs{x}_{n+1}^{\text{SPNN}}||_2^2.
\end{equation}

\item \textbf{Fulfillment of the degeneracy conditions}: The loss function will also account for the degeneracy conditions, Eq. (\ref{eq:degendisc}) in order to ensure the thermodynamic consistency of the solution, implemented as the sum of the squared elements of the degeneracy vectors for each time step,
\begin{equation}\label{eq:loss_degen}
\mathcal{L}_n^{\text{degen}} = ||\mathsf{L}_n\cdot\mathsf{DS}_n||_2^2 + ||\mathsf{M}_n\cdot\mathsf{DE}_n||_2^2.
\end{equation}
This term acts as a regularization of the loss function and, at the same time, is the responsible of ensuring thermodynamic consistency of the integration scheme. This is, in other words, our inductive bias.

\item \textbf{Regularization}: In order to avoid overfitting, an extra L2 regularization term $\mathcal{L}^{\text{reg}}$ is added to the loss function,
\begin{equation}\label{eq:loss_regular}
\mathcal{L}^{\text{reg}}=\sum_l^{L}\sum_i^{n^{[l]}}\sum_j^{n^{[l+1]}}{(w_{i,j}^{[l],\text{SPNN}})^2}.
\end{equation}

\end{itemize}

The same database split procedure is followed as in the SAE, dividing the complete dataset of $N_T$ snapshots in a partition of train snapshots ($N_{\text{train}}=80\%$ of the database snapshots) and test snapshots ($N_{\text{test}}=20\%$ of the database snapshots) so that $N_T=N_{\text{train}}+N_{\text{test}}$. The total loss function is computed as the mean squared error (MSE) of the data loss and degeneracy residual, in addition to the regularization term, for all the training snapshots ($N_{\text{train}}$) of the simulation time $T$. Both the data loss error and the regularization terms are weighted with two additional hyperparameters $\lambda_d^{\text{SPNN}}$ and $\lambda_r^{\text{SPNN}}$ respectively, which account for their relative influence in the total loss function with respect to the degeneracy constraint,
\begin{equation}\label{eq:loss_total}
\mathcal{L}^{\text{SPNN}} = \frac{1}{N_{\text{train}}}\sum_{n=0}^{N_{\text{train}}}{(\lambda_d^{\text{SPNN}}\mathcal{L}_n^{\text{data}} + \mathcal{L}_n^{\text{degen}})} + \lambda_r^{\text{SPNN}}\mathcal{L}^{\text{reg}}.
\end{equation}

The usual backpropagation algorithm \cite{paszke2017automatic} is then used to calculate the gradient of the loss function for each net parameter (weight and bias vectors), which are updated with the gradient descent technique \cite{ruder2016overview}. The training algorithm is sketched below in \myalgref{alg:SPNN_train}.

\begin{algorithm*}[h!]
\caption{Pseudocode for the train algorithm of the SPNN.}\label{alg:SPNN_train}
\begin{algorithmic}
  \State \textbf{Load train database:} $\bs{z}^{\text{SAE}}$ (train partition), $\Delta t$;
  \State \textbf{Define network architecture:} $N_{\text{in}}^{\text{SPNN}}=d$, $N_{\text{out}}^{\text{SPNN}}=d\cdot(d+3)$, $N_{h}^{\text{SPNN}}$, $\sigma_j^{\text{SPNN}}$;
  \State \textbf{Define hyperparameters:} $l_r^{\text{SPNN}}$, $\lambda_d^{\text{SPNN}}$, $\lambda_r^{\text{SPNN}}$;
  \State Initialize $w_{i,j}^{\text{SPNN}}$, $b_j^{\text{SPNN}}$;
  \For{each epoch}
  		\State Initialize loss function: $C=0$;
  		\For{each train snapshot}
  			\State Encoder: $\bs{x}^{\text{SAE}}_n=q_\phi(\bs{z}^{\text{GT}}_n)$; \Comment \myeqref{eq:decoder} 
  			\State Forward propagation: $[\mathsf{L}_n$, $\mathsf{M}_n$, $\mathsf{DE}_n$, $\mathsf{DS}_n] \gets \text{SPNN}(\bs{x}^{\text{SAE}}_n)$; \Comment \myeqref{eq:forward_net}
  			\State Time step integration: $\bs{x}^{\text{SPNN}}_{n+1} \gets \bs{x}^{\text{SAE}}_n + \Delta t\;(\mathsf{L}_n\cdot\mathsf{DE}_n + \mathsf{M}_n\cdot\mathsf{DE}_n)$; \Comment \myeqref{eq:generic_disc}
  			\State Update loss function: $C \gets C + \lambda_d^{\text{SPNN}}\mathcal{L}_n^{\text{data}} + \mathcal{L}_n^{\text{degen}}$; \Comment \myeqref{eq:loss_data}, \myeqref{eq:loss_degen}
  		\EndFor		
  		\State MSE loss function: $\mathcal{L}^{\text{SPNN}} \gets \frac{C}{N_{\text{train}}} + \lambda_r^{\text{SPNN}}\mathcal{L}^{\text{reg}}$ \Comment \myeqref{eq:loss_regular}, \myeqref{eq:loss_total}
  		\State Backward propagation;
  		\State Optimizer step;
  \EndFor
\end{algorithmic}
\end{algorithm*}

The testing consists of the full time integration of the initial state vector $\bs{z}_0$ at $t=0$ along the complete simulation time interval $\mathcal{I}=(0,T]$, reproducing the problem statement established in \myeqref{eq:pde}. Thus, the net performance is evaluated with the mean squared error (MSE) of the SPNN state variable predictions and the ground truth solution for the complete set of snapshots $N_T$,
\begin{equation}\label{eq:SPNN_error}
\text{MSE}^{\text{SPNN}}\;(\bs{z})=\frac{1}{N_T}\sum_{n=0}^{N_T}\varepsilon_n^2=\frac{1}{N_T}\sum_{n=0}^{N_T} \left(\bs{x}^{\text{GT}}_{n}-\bs{z}^{\text{SPNN}}_{n}\right)^2,
\end{equation}
tested for the same nonlinear systems trained in the SAE training phase. A pseudocode of the testing process of the SPNN is shown in \myalgref{alg:SPNN_test}.

\begin{algorithm*}[h!]
\caption{Pseudocode for the test algorithm of the complete integration scheme of the SPNN.}\label{alg:SPNN_test}
\begin{algorithmic}
  \State \textbf{Load database:} $\bs{z}^{\text{GT}}$, $\Delta t$;
  \State \textbf{Load network parameters};
  \State Initialize state vector: $\bs{z}_0^{\text{SAE}}=\bs{z}_0^{\text{SPNN}}=\bs{z}^{\text{GT}}_0$;
  \State Initialize encoded state vector: $\bs{x}_0^{\text{SAE}}=\bs{x}_0^{\text{SPNN}}=q_\phi(\bs{z}^{\text{GT}}_0)$; \Comment \myeqref{eq:encoder}
  \For{each snapshot}   
  	\State Forward propagation: $[\mathsf{L}_n$, $\mathsf{M}_n$, $\mathsf{DE}_n$, $\mathsf{DS}_n] \gets \text{SPNN}(\bs{x}^{\text{SPNN}}_n)$; \Comment \myeqref{eq:forward_net}
  	\State Time step integration: $\bs{x}^{\text{SPNN}}_{n+1} \gets \bs{x}^{\text{SPNN}}_n + \Delta t\;(\mathsf{L}_n\cdot\mathsf{DE}_n + \mathsf{M}_n\cdot\mathsf{DE}_n)$; \Comment \myeqref{eq:generic_disc}
  	\State Update state vector: $\bs{x}_{n}^{\text{SPNN}} \gets \bs{x}^{\text{SPNN}}_{n+1}$;
  	\State Update snapshot: $n \gets n+1$;
  	\State Decoder: $\bs{z}^{\text{SPNN}}_{n+1}=p_\theta(\bs{x}^{\text{SPNN}}_{n+1})$; \Comment \myeqref{eq:decoder} 
  	\State Compute Squared Error: $\varepsilon_{n+1}^2=\left(\bs{z}_{n+1}^{\text{GT}}-\bs{z}_{n+1}^{\text{SPNN}}\right)^2$; \Comment \myeqref{eq:SPNN_error}		
  \EndFor	
  \State Compute $\text{MSE}^{\text{SPNN}}\;(\bs{z})$; \Comment \myeqref{eq:SPNN_error}	
\end{algorithmic}
\end{algorithm*}

{The SPNN is compared on each example with a baseline unconstrained neural network which directly predicts the time evolution of the latent vector $x_{t+1}$ from the current snapshot $x_t$, with a similar training and integration scheme as depicted in \myalgref{alg:SPNN_train} and \myalgref{alg:SPNN_test}.}

\section{Validation examples: Couette flow of an Oldroyd-B fluid}\label{sec:visco}

\subsection{Description}

The first example is a shear (Couette) flow of an Oldroyd-B fluid model. This is a constitutive model for viscoelastic fluids. It arises from the consideration of linear elastic dumbbells as a proxy representation of polymeric chains immersed in a solvent.

\begin{figure}[h]
\centering
   \includegraphics{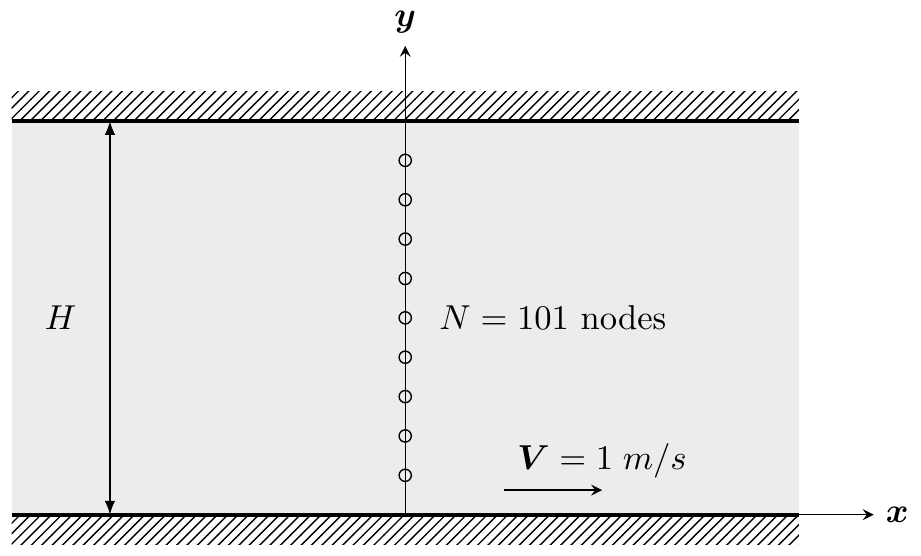}  
\caption{Couette flow in an Olroyd-B fluid. Horizontal position, velocity, internal energy and conformation tensor shear component are tracked for the total of 100 nodes (excluded the $y=H$ node).\label{fig:visco}}
\end{figure}

The problem is solved by the CONNFFESSIT technique \cite{laso1993calculation}, based on the Fokker-Plank equation \cite{le2009multiscale}. This equation is solved by converting it in its corresponding It\^o stochastic differential equation,
\begin{alignat}{1}\label{eq:visco_SDE}
 dr_x &=\left(\dpar{u}{y}r_y-\frac{1}{2\text{We}}r_x\right)dt+\frac{1}{\sqrt{\text{We}}}dV_t, \nonumber\\
 dr_y &=-\frac{1}{2\text{We}}r_ydt+\frac{1}{\sqrt{\text{We}}}dW_t,
\end{alignat}
where $\bs{r}=[r_x,\;r_y]^\top$, $r_x=r_x(y,t)$ and assuming a Couette flow so that $r_y=r_y(t)$ depends only on time, ``We" stands for the Weissenberg number and $V_t$, $W_t$ are two independent one-dimensional Brownian motions. This equation is solved via Monte Carlo techniques, by replacing the mathematical expectation by the empirical mean.

The model relies on the microscopic description of the state of the dumbbells. Thus, it is particularly useful to base the microscopic description on the evolution of the conformation tensor $\bs c=\langle\bs{r}\bs{r}\rangle$, this is, the second moment of the dumbbell end-to-end distance distribution function. This tensor is in general not experimentally measurable and plays the role of an internal variable. The expected $xy$ stress component tensor will be given by
\begin{equation*}\label{eq:visco_tau}
\tau=\frac{\epsilon}{\text{We}}\frac{1}{K}\sum_{k=1}^Kr_xr_y,
\end{equation*}
where $K$ is the number of simulated dumbbells and $\epsilon=\frac{\nu_p}{\nu_p}$ is the ratio of the polymer to solvent viscosities.

The state variables chosen for the full order model are the position of the fluid on each node of the mesh $\bs{q}$, see Fig. \ref{fig:visco}, its velocity $\bs{v}$ in the x direction, internal energy $e$ and the conformation tensor shear component $\tau$ for all the nodes of the mesh,
\begin{equation}\label{eq:visco_z}
\mathcal{S}=\{\bs{z}=(q_i,v_i,e_i,\tau_i,i=1,2,...,N)\in(\mathbb{R}\times\mathbb{R}\times\mathbb{R}\times\mathbb{R})^N\},
\end{equation}
resulting in a full-order model of $D=4\cdot N$ dimensions.

\subsection{Database and Hyperparameters}

The training database for this Olroyd-B model is generated in MATLAB with a multiscale approach \cite{le2009multiscale} in dimensionless form. The fluid is discretized in the vertical direction with $N=100$ elements (101 nodes) in a total height of $H=1$. A total of 10,000 dumbells were considered at each nodal location in the model. The lid velocity is set to $V=1$, the viscolastic Weissenberg number We $=1$ and Reynolds number of Re $ = 0.1$. The simulation time of the movement is $T = 1$ in time increments of $\Delta t = 0.0067$ ($N_T=150$ snapshots). 

The database consists of the state vector, \myeqref{eq:visco_z}, of the 100 nodal trajectories (excluding the node at $y=H$, for which a no-slip condition $v=0$ has been imposed) for each snapshot of the simulation. This database is split in 120 train snapshots and 30 test snapshots. 

The SAE input and output sizes are $N_{\text{in}}^{\text{SAE}} = N_{\text{out}}^{\text{SAE}} = D = 4\cdot N = 400$. The number of hidden layers in both the encoder and decoder is $N_h^{\text{SAE}} = 2$ with 160 neurons each, ReLU activation functions and linear in the first and last layer. The number of bottleneck variables is set to $N_d=10$. It is initialized according to the Kaiming method \cite{he2015delving}, with normal distribution and the optimizer used is Adam \cite{kingma2014adam}, with a learning rate of $l_r^{\text{SAE}}=10^{-4}$. The sparsity parameter is set to $\lambda_r^{\text{SAE}}=10^{-4}$. The training process (\myalgref{alg:SAE_train}) is able to sparsify the bottleneck variables of the Olroyd-B model with only $d=4$ latent variables, which are the input variables used in the structure preserving-neural network.

Thus, the SPNN input and output size are $N_{\text{in}}^{\text{SPNN}} = d = 4$ and $N_{\text{out}}^{\text{SPNN}} = d\cdot(d+2) = 24$. The number of hidden layers is $N_h^{\text{SPNN}} = 5$ with 24 neurons each, ReLU activation functions and linear in the last layer. The same initialization method and optimizer are used as in the SAE network, with a learning rate of $l_r^{\text{SPNN}}=10^{-5}$. The weight decay and the data weight are set to $\lambda_r^{\text{SPNN}}=10^{-5}$ and $\lambda_d^{\text{SPNN}}=10^3$ respectively.

{The unconstrained network training parameters are analogous to the structure-preserving network, except for the output size $N_{out}^{UC} = N_{in}^{UC} = 4$. Several network architectures were tested, and the lowest error is achieved with $N_h = 5$ hidden layers and 25 neurons each layer.}

\subsection{Results}

\begin{figure*}[h!]

\centering
\begin{tikzpicture}[scale=0.9]

% Latent Variables
\begin{axis}[name=plot1,
  grid=major, % Display a grid
  grid style={dashed,gray!30}, % Set the style
  xlabel={$t$ [s]},
  ylabel={$\bs{x}$ [-]},
  cycle list name = {color list}]
  
  \foreach \F in {1,...,10}{
	\addplot table [y=var_\F, x=tspan]{graphs/visco/latent_AE.txt};
	}
\end{axis}

% SPNN
\begin{axis}[name=plot2, at={($(plot1.right of south east)+(2cm,0)$)},
  grid=major, % Display a grid
  grid style={dashed,gray!30}, % Set the style
  xlabel={$t$ [s]},
  ylabel={$\bs{x}$ [-]},
  cycle list name = {color list}]
  
  \foreach \F in {1,2,3,4}{
	\addplot [color=black, dashed] table [y=x_AE_\F, x=tspan]{graphs/visco/results_SPNN.txt};
	\ifthenelse{\F=1}{\addlegendentry{SAE}}{}
	\addplot [color=blue] table [y=x_SPNN_\F, x=tspan]{graphs/visco/results_SPNN.txt};
	\ifthenelse{\F=1}{\addlegendentry{SPNN}}{}
	}
\end{axis}
	
% Thermodynamics
\begin{axis}[name=plot3, at={($(plot1.below south west)+(0,-6cm)$)},
  grid=major, % Display a grid
  grid style={dashed,gray!30}, % Set the style
  xlabel={$t$ [s]},
  ylabel={$dE/dt, dS/dt$ [-]}]
  
	\addplot [color=red] table [y=dEdt, x=tspan]{graphs/visco/viscolastic_thermodynamics.txt};
	\addlegendentry{dE/dt}
	\addplot [color=blue] table [y=dSdt, x=tspan]{graphs/visco/viscolastic_thermodynamics.txt};
	\addlegendentry{dS/dt}
\end{axis}

\end{tikzpicture}
\caption{Top Left: Time evolution of the latent variables encoded with the sparse autoencoder (SAE) in the Olroyd-B fluid problem. The bottleneck has $N_d=10$ neurons and the learning algorithm automatically sparsifies them to a dimensionality of $d=4$ relevant latent variables. Top Right: Time evolution of the relevant latent variables integrated in time by the structure-preserving neural network (SPNN). Bottom: Evolution of the time derivative of the energy ($dE/dt$) and entropy ($dS/dt$) of the latent variables.}
\label{fig:visco_results_latent}
\end{figure*}
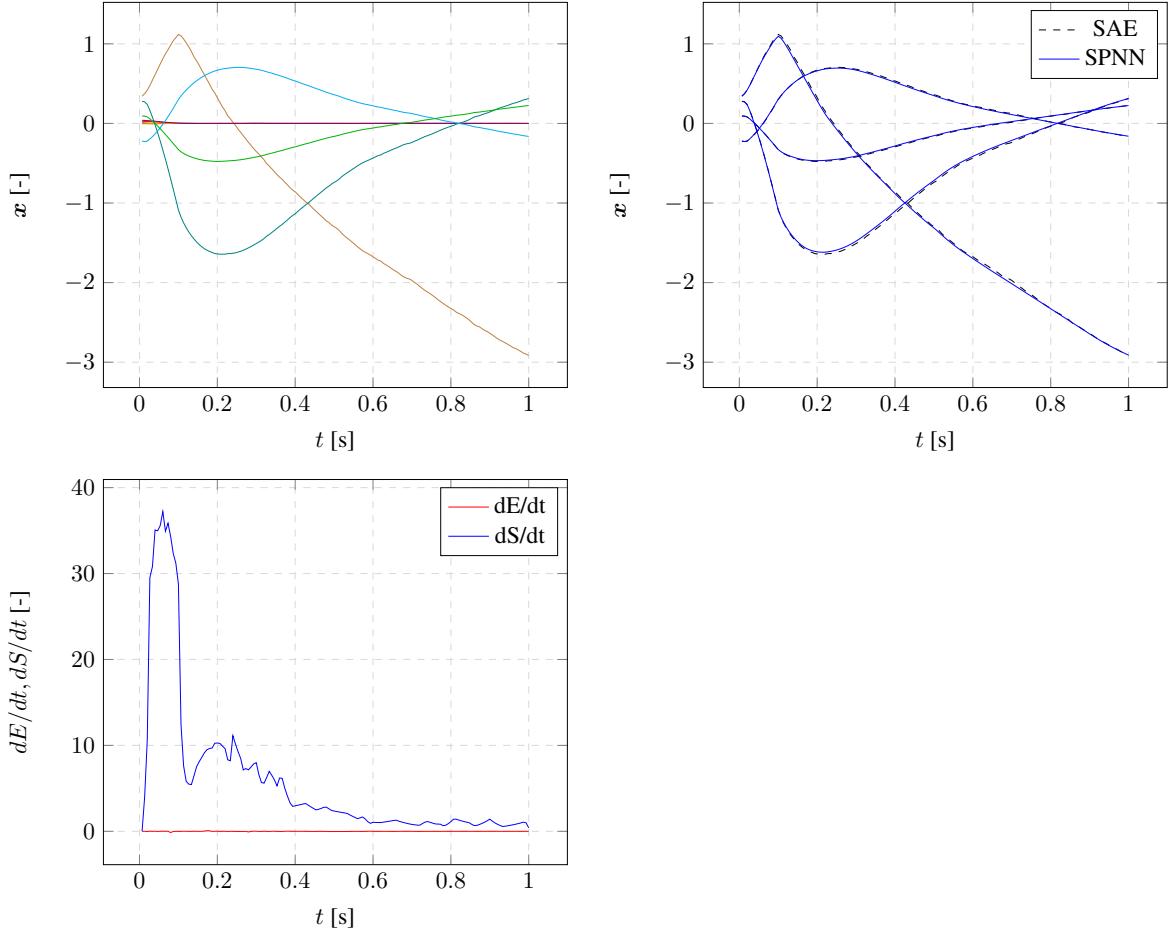

\myfigref{fig:visco_results_latent} shows the time evolution of the SAE bottleneck variables after the complete training process. The sparsity constraint forces the unnecessary latent variables to vanish, remaining a learnt latent dimensionality of $d=4$ relevant variables from a starting bottleneck dimension of $N_d = 10$ (\myfigref{fig:visco_results_latent}, {Top} Left). This compares advantageously with the obtained dimensionality $d = 6$ of our previous work \cite{gonzalez2019thermodynamically}. \mytabref{tab:visco_results_SAE} shows the mean squared error of the SAE reconstruction, computed with \myalgref{alg:SAE_test}, and an equal reduction using Proper Orthogonal Decomposition. Then, the SPNN is able to integrate the whole trajectory of the relevant latent variables in the reduced manifold in good agreement with the original SAE reduction (\myfigref{fig:visco_results_latent}, {Top} Right). {The integration scheme also ensures that the time derivative of the energy ($dE/dt$) and entropy ($dS/dt$) of the system remain equal to zero or greater than zero respectively, in fulfilment with the first and second law of thermodynamics (\myfigref{fig:visco_results_latent}, Bottom Left), computed with \myeqref{eq:Econs}.}

\begin{table}[h]
\caption{Left: Mean squared error of the SAE reconstruction ($\text{MSE}^{\text{SAE}}$) for the 4 state variables of the Olroyd-B Couette flow example, reported only for the test snapshots. Right: Mean squared error of the same reduction using a Proper Orthogonal Decomposition algorithm ($\text{MSE}^{\text{POD}}$).}\label{tab:visco_results_SAE}
\centering
\begin{tabular}{lll}
\hline
State variable ($\bs{z}_i$) & $\text{MSE}^{\text{SAE}}$ & $\text{MSE}^{\text{POD}}$ \\ \hline
$q$ [-]                     & $2.52\cdot10^{-6}$ & $7.87\cdot10^{-6}$ \\ \hline
$v$ [-]                     & $7.27\cdot10^{-5}$ & $4.31\cdot10^{-5}$ \\ \hline
$e$ [-]                     & $1.89\cdot10^{-6}$ & $7.33\cdot10^{-6}$ \\ \hline
$\tau$ [-]                  & $7.22\cdot10^{-6}$ & $2.07\cdot10^{-5}$ \\ \hline
\end{tabular}
\end{table}

\begin{figure*}[h!]
\centering
\begin{tikzpicture}[scale=0.9]

% q
\begin{axis}[name=plot1,
  grid=major, % Display a grid
  grid style={dashed,gray!30}, % Set the style
  xlabel={$t$ [s]},
  ylabel={$q$ [-]},
  cycle list name = {color list},
  legend pos=north west]
  
  \foreach \F in {20,40,60,80}{
	\addplot [color=black, dashed] table [y=GT_\F, x=tspan]{graphs/visco/results_q.txt};
	\ifthenelse{\F=20}{\addlegendentry{GT}}{}
	\addplot [color=blue] table [y=SPNN_\F, x=tspan]{graphs/visco/results_q.txt};
	\ifthenelse{\F=20}{\addlegendentry{SPNN}}{}
	}
\end{axis}

% vx
\begin{axis}[name=plot2, at={($(plot1.right of south east)+(2cm,0)$)},
  grid=major, % Display a grid
  grid style={dashed,gray!30}, % Set the style
  xlabel={$t$ [s]},
  ylabel={$v$ [-]},
  cycle list name = {color list}, 
  ymax = 1.1]
  
  \foreach \F in {20,40,60,80}{
	\addplot [color=black, dashed] table [y=GT_\F, x=tspan]{graphs/visco/results_vx.txt};
	\ifthenelse{\F=20}{\addlegendentry{GT}}{}
	\addplot [color=blue] table [y=SPNN_\F, x=tspan]{graphs/visco/results_vx.txt};
	\ifthenelse{\F=20}{\addlegendentry{SPNN}}{}
	}	
\end{axis}
	
% e
\begin{axis}[name=plot3, at={($(plot1.below south west)+(0,-6cm)$)},
  grid=major, % Display a grid
  grid style={dashed,gray!30}, % Set the style
  xlabel={$t$ [s]},
  ylabel={$e$ [-]},
  cycle list name = {color list},
  legend pos=north west]
  
  \foreach \F in {20,40,60,80}{
	\addplot [color=black, dashed] table [y=GT_\F, x=tspan]{graphs/visco/results_e.txt};
	\ifthenelse{\F=20}{\addlegendentry{GT}}{}
    \addplot [color=blue] table [y=SPNN_\F, x=tspan]{graphs/visco/results_e.txt};
	\ifthenelse{\F=20}{\addlegendentry{SPNN}}{}
	}	
\end{axis}

% tau
\begin{axis}[name=plot4, at={($(plot3.right of south east)+(2cm,0)$)},
  grid=major, % Display a grid
  grid style={dashed,gray!30}, % Set the style
  xlabel={$t$ [s]},
  ylabel={$\tau$ [-]},
  cycle list name = {color list}]
  
  \foreach \F in {20,40,60,80}{
	\addplot [color=black, dashed] table [y=GT_\F, x=tspan]{graphs/visco/results_tau.txt};
	\ifthenelse{\F=20}{\addlegendentry{GT}}{}
	\addplot [color=blue] table [y=SPNN_\F, x=tspan]{graphs/visco/results_tau.txt};
	\ifthenelse{\F=20}{\addlegendentry{SPNN}}{}
	}	
\end{axis}

\end{tikzpicture}
\caption{Results of the complete integration scheme (SPNN) with respect to the ground truth simulation (GT) for 4 different nodes of the Olroyd-B fluid database.}
\label{fig:visco_results}
\end{figure*}
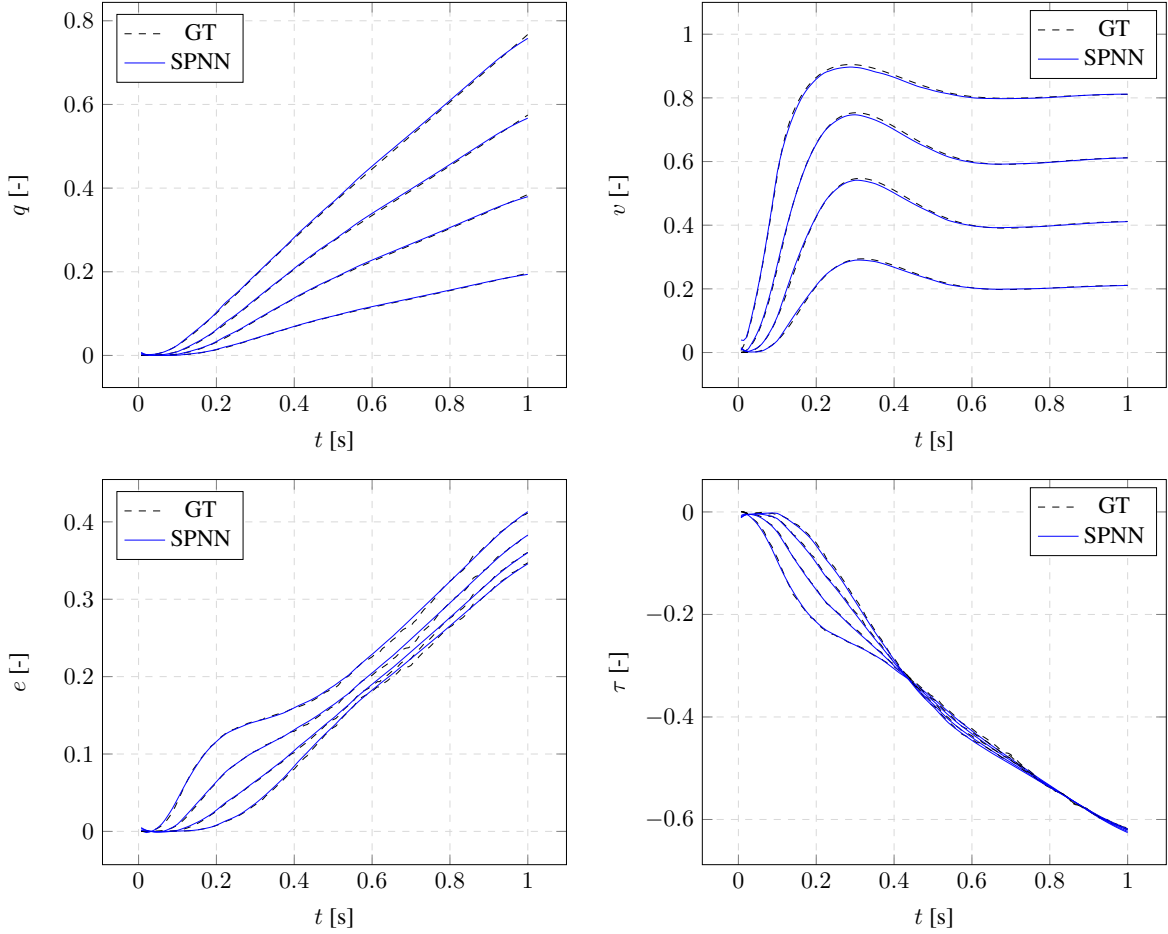

\myfigref{fig:visco_results} presents the time evolution of the decoded state variables of the Olroyd-B Couette flow for 4 different nodes computed with the presented integration scheme and the ground truth. The results show a good agreement in the transient response of the Couette flow, even for the high nonlinearities of the internal energy and the conformation tensor shear component. The mean squared error of the total integration scheme, computed with \myalgref{alg:SPNN_test}, for the 4 state variables is reported in \mytabref{tab:visco_results_SPNN} { using the Structure-preserving neural network (SPNN) and the unconstrained approach (UC). Our neural network achieves less error than the unconstrained one, showing the importance of adding the physical constrains to the learning process, and this difference becomes greater in the second example}.

\begin{table}[h!]

\caption{Mean squared error of the SPNN integration scheme and the unconstrained (UC) approach for the 4 state variables of the Olroyd-B Couette flow example, reported for the complete trajectory.}\label{tab:visco_results_SPNN}
\centering
\begin{tabular}{lcc}
\hline
State variable ($\bs{z}_i$) & $\text{MSE}^{\text{SPNN}}$ & $\text{MSE}^{\text{UC}}$  \\ \hline
$q$ [-]                     & $1.78\cdot10^{-5}$ & $7.96\cdot10^{-5}$ \\ \hline
$v$ [-]                     & $3.34\cdot10^{-5}$ & $3.48\cdot10^{-5}$ \\ \hline
$e$ [-]                     & $5.60\cdot10^{-6}$ & $5.67\cdot10^{-5}$ \\ \hline
$\tau$ [-]                  & $2.19\cdot10^{-5}$ & $1.22\cdot10^{-4}$ \\ \hline
\end{tabular}
\end{table}

\section{Rolling Hyperelastic Tire}\label{sec:tire}

\subsection{Description}

\begin{figure}[h]
\centering
   \includegraphics[width=0.85\textwidth]{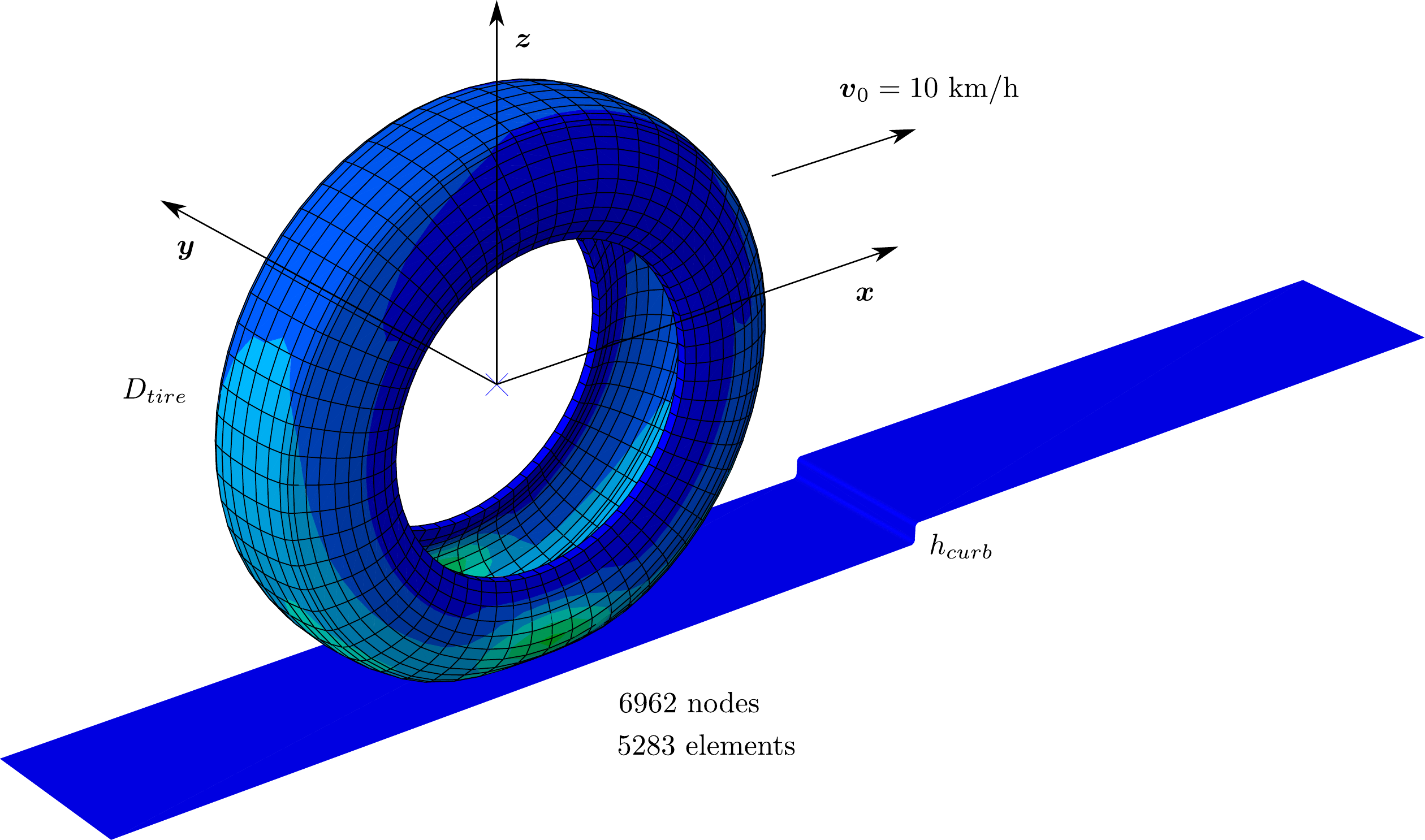}  
\caption{Hyperelastic tire rolling towards a curb. 3D position, 3D velocity and Cauchy stress tensor components are tracked for the total of 4140 selected nodes.\label{fig:tire}}
\end{figure}

The second example is a simulation of the transient response of a 175 SR14 rolling tire ($D_{\text{tire}}=0.66$ m) impacting with a curb ($h_{\text{curb}}=0.025$ m). The tire is initially preloaded with an inflation load of 200 kPa, simulating the internal air pressure, and a footprint load of 3300 N in the vertical axis, simulating the weight of the vehicle distributed equally in the tires. The free rolling conditions are determined in a separated analysis, corresponding to $\omega=8.98$ rad/s for a translational horizontal velocity of $v_0=10$ km/h (see \myfigref{fig:tire}). 

The tread and sidewalls of the tire are made of rubber, modeled as an incompressible hyperelastic material with a viscolastic component described by a one-term Prony series of the dimensionless shear relaxation modulus,
$$
g_R(t)=1-\bar{g}_1(1-e^{\frac{-t}{\tau_1}}),
$$
with relaxation coefficient of $\bar{g}_1=0.3$ and relaxation time of $\tau_1=0.1$ s. The belts and carcass of the tire are constructed from fiber-reinforced rubber composites, modeled as a linear elastic material, with a \ang{20} orientation of the reinforcing belt.

The state variables chosen for the full order model are the 3D position $\bs{q}_i$, velocity $\bs{v}_i$ and the 6 different components of the Cauchy stress tensor $\bs{\sigma}_i$ for each $i$ node of the studied mesh subset $N$,
\begin{equation}\label{eq:tire_z}
\mathcal{S}=\{\bs{z}=(\bs{q}_i,\bs{v}_i,\bs{\sigma}_i,i=1,2,...,N)\in(\mathbb{R}^3\times\mathbb{R}^3\times\mathbb{R}^6)^N\},
\end{equation}
resulting in a full-order model of $D=12\cdot N$ dimensions.

\subsection{Database and Hyperparameters}

The training database for this rolling tire simulation is generated by finite element simulation. The full-order model is discretized with 5283 elements in a total of 6962 nodes. The simulation time of the movement is $T = 0.5$ s in time increments of $\Delta t = 0.0025$ s ($N_T=200$ snapshots). The database consists of the normalized state vector (\myeqref{eq:tire_z}) of a subset of $N=4140$ relevant nodes in every time step snapshot. The total state vector snapshots are randomly split in 160 train snapshots and 40 test snapshots. 

The SAE architecture for this second example is slightly modified in order to handle the high dimensionality of the problem. The three physical variables ($\bs{q}$, $\bs{v}$, and $\bs{\sigma}$) are encoded and decoded independently, due to their very different nature. In this way, three bottleneck latent vectors are obtained. The input and output sizes of the three SAEs are $N_{\text{in},q}^{\text{SAE}} = N_{\text{out},q}^{\text{SAE}} = 3\cdot N = 12420$ for the position variable, $N_{\text{in},v}^{\text{SAE}} = N_{\text{out},v}^{\text{SAE}} = 3\cdot N = 12420$ and $N_{\text{in},\sigma}^{\text{SAE}} = N_{\text{out},\sigma}^{\text{SAE}} = 6\cdot N = 24840$ for the stress tensor. 

The number of hidden layers in both the encoder and decoder is $N_{h}^{\text{SAE}} = 2$ in the three variables with 40 neurons each in position and velocity, and 80 neurons in the stress tensor, with ReLU activation functions and linear in the first and last layers. The number of bottleneck variables is set to $N_{d,q}=10$ for the position, $N_{d,v}=10$  for velocity and $N_{d,\sigma}=20$  for the stress tensor. Thus, the total dimensionality of the bottleneck latent vector is $N_d=N_{d,q}+N_{d,v}+N_{d,\sigma}=40$.

In the same way as we do in the first example, the nets are initialized according to the Kaiming method \cite{he2015delving}, with normal distribution and the optimizer used is Adam \cite{kingma2014adam}, with a learning rate of $l_r^{\text{SAE}}=10^{-4}$. The sparsity parameter, in this case, is set to $\lambda_r^{\text{SAE}}=10^{-2}$. The training process (\myalgref{alg:SAE_train}) is able to sparsify the bottleneck variables of the rolling tire model with only $d_q=4$ position, $d_q=3$ velocity and $d_\sigma=2$ stress tensor latent variables. So, the learnt dimensionality of the reduced model is $d=d_q+d_v+d_\sigma=9$, which are the input variables used in the structure preserving-neural network.

Thus, the SPNN input and output sizes are $N_{\text{in}}^{\text{SPNN}} = d =9$ and $N_{\text{out}}^{\text{SPNN}} = d\cdot(d+2) = 99$. The number of hidden layers is $N_h^{\text{SPNN}} = 5$ with 198 neurons each, with ReLU activation functions and linear in the last layer. The same initialization method and optimizer are used as in the SAE network, with a learning rate of $l_r^{\text{SPNN}}=10^{-5}$. The weight decay and the data weight are set to $\lambda_r^{\text{SPNN}}=10^{-4}$ and $\lambda_d^{\text{SPNN}}=10^3$ respectively.

{The unconstrained network training parameters are analogous to the structure-preserving network, except for the output size $N_{out}^{UC} = N_{in}^{UC} = 9$. Several network architectures were tested, and the lowest error is achieved with $N_h = 5$ hidden layers and 45 neurons each layer.}

\subsection{Results}

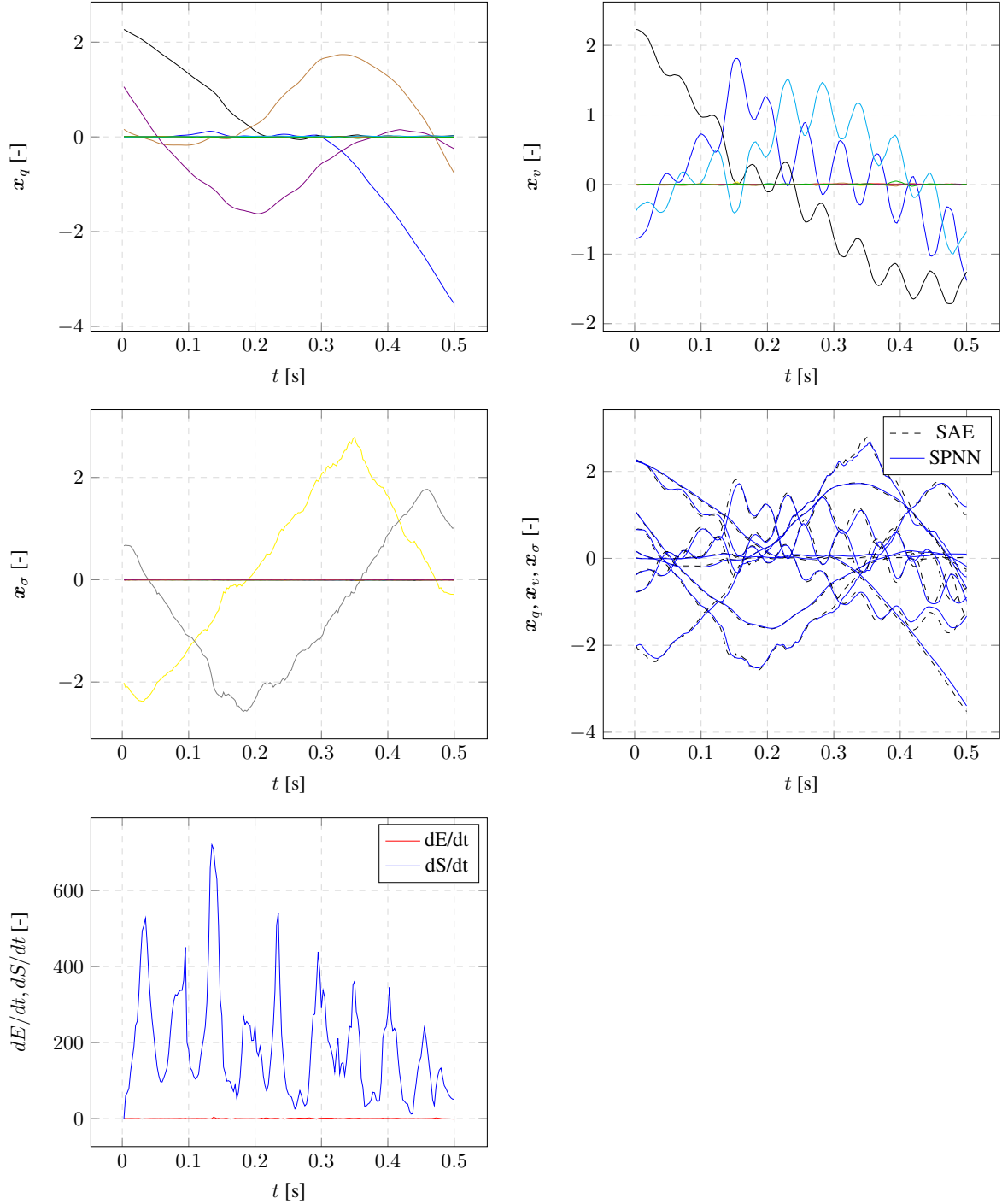
\begin{figure*}[h!]

\centering
\begin{tikzpicture}[scale=0.9]

% Latent q
\begin{axis}[name=plot1,
  grid=major, % Display a grid
  grid style={dashed,gray!30}, % Set the style
  xlabel={$t$ [s]},
  ylabel={$\bs{x}_q$ [-]},
  cycle list name = {color list}]
  
  \foreach \F in {1,...,10}{
	\addplot table [y=var_\F, x=tspan]{graphs/tire/latent_AE_q.txt};
	}
\end{axis}

% Latent v
\begin{axis}[name=plot2, at={($(plot1.right of south east)+(2cm,0)$)},
  grid=major, % Display a grid
  grid style={dashed,gray!30}, % Set the style
  xlabel={$t$ [s]},
  ylabel={$\bs{x}_v$ [-]},
  cycle list name = {color list}]
  
  \foreach \F in {1,...,10}{
	\addplot table [y=var_\F, x=tspan]{graphs/tire/latent_AE_v.txt};
	}	
\end{axis}
	
% Latent sigma
\begin{axis}[name=plot3, at={($(plot1.below south west)+(0,-6cm)$)},
  grid=major, % Display a grid
  grid style={dashed,gray!30}, % Set the style
  xlabel={$t$ [s]},
  ylabel={$\bs{x}_\sigma$ [-]},
  cycle list name = {color list}]
  
  \foreach \F in {1,...,20}{
	\addplot table [y=var_\F, x=tspan]{graphs/tire/latent_AE_sigma.txt};
	}	
\end{axis}

% SPNN
\begin{axis}[name=plot4, at={($(plot3.right of south east)+(2cm,0)$)},
  grid=major, % Display a grid
  grid style={dashed,gray!30}, % Set the style
  xlabel={$t$ [s]},
  ylabel={$\bs{x}_q$, $\bs{x}_v$, $\bs{x}_\sigma$ [-]},
  cycle list name = {color list}]
  
    \foreach \F in {1,...,9}{
	\addplot [color=black, dashed] table [y=x_AE_\F, x=tspan]{graphs/tire/results_SPNN.txt};
	\ifthenelse{\F=1}{\addlegendentry{SAE}}{}
	\addplot [color=blue] table [y=x_SPNN_\F, x=tspan]{graphs/tire/results_SPNN.txt};
	\ifthenelse{\F=1}{\addlegendentry{SPNN}}{}
	}	
\end{axis}

% Thermodynamics
\begin{axis}[name=plot3, at={($(plot3.below south west)+(0,-6cm)$)},
  grid=major, % Display a grid
  grid style={dashed,gray!30}, % Set the style
  xlabel={$t$ [s]},
  ylabel={$dE/dt, dS/dt$ [-]}]
  
	\addplot [color=red] table [y=dEdt, x=tspan]{graphs/tire/rolling_tire_thermodynamics.txt};
	\addlegendentry{dE/dt}
	\addplot [color=blue] table [y=dSdt, x=tspan]{graphs/tire/rolling_tire_thermodynamics.txt};
	\addlegendentry{dS/dt}
\end{axis}

\end{tikzpicture}
\caption{Time evolution of the latent variables encoded with the sparse autoencoder (SAE) in the hyperelastic rolling tire problem. The bottleneck has $N_d=40$ neurons and the learning algorithm sparsifies them to a dimensionality of $d=9$ relevant latent variables. {{Bottom} Middle} Right: Time evolution of the relevant latent variables integrated in time by the structure-preserving neural network (SPNN). {Bottom: Evolution of the time derivative of the energy ($dE/dt$) and entropy ($dS/dt$) of the latent variables.}}
\label{fig:tire_results_latent}
\end{figure*}

\myfigref{fig:tire_results_latent} shows the time evolution of the SAE bottleneck variables ($\bs{x}_q$, $\bs{x}_v$ and $\bs{x}_\sigma$) after the complete training process. The sparsity constraint forces the unnecessary latent variables to vanish, remaining a learnt latent dimensionality of $d_q=4$, $d_v=3$ and $d_\sigma=2$ relevant variables from a starting bottleneck dimension of $N_{d,q} = 10$, $N_{d,v} = 10$ and $N_{d,\sigma} = 20$ respectively (\myfigref{fig:tire_results_latent}). The mean squared error of the SAE reconstruction, computed with \myalgref{alg:SAE_test}, and a equal reduction with a Proper Orthogonal Decomposition is shown in \mytabref{tab:tire_results_SAE}. Then, the SPNN is able to integrate the whole trajectory of the relevant latent variables in the reduced manifold in good agreement with the original SAE reduction (\myfigref{fig:tire_results_latent}, {{Bottom} Middle} Right). {Also, the integration scheme fulfils the first and second laws of thermodynamics (\myfigref{fig:tire_results_latent}, Bottom) computed with \myeqref{eq:Econs}}

\begin{table}[h]
\caption{Left: Mean squared error of the SAE reconstruction ($\text{MSE}^{\text{SAE}}$) for the 12 state variables of the rolling tire example, reported only for the test snapshots. Right: Mean squared error of the same reduction using a Proper Orthogonal Decomposition algorithm ($\text{MSE}^{\text{POD}}$).}\label{tab:tire_results_SAE}
\centering
\begin{tabular}{lll}
\hline
State variable ($\bs{z}_i$) & $\text{MSE}^{\text{SAE}}$ & $\text{MSE}^{\text{POD}}$ \\ \hline
$q_1$ [m]         	     & $2.37\cdot10^{-5}$ & $1.30\cdot10^{-3}$ \\ 
$q_2$ [m]                 & $3.69\cdot10^{-7}$ & $6.27\cdot10^{-7}$ \\ 
$q_3$ [m]                 & $3.06\cdot10^{-5}$ & $6.55\cdot10^{-5}$ \\ \hline
$v_1$ [m/s]               & $1.00\cdot10^{-3}$ & $3.32\cdot10^{-2}$ \\ 
$v_2$ [m/s]               & $4.54\cdot10^{-5}$ & $2.37\cdot10^{-2}$ \\ 
$v_3$ [m/s]               & $3.70\cdot10^{-3}$ & $6.91\cdot10^{-2}$ \\ \hline
$\sigma_{11}$ [MPa]       & $2.41\cdot10^{-4}$ & $3.74\cdot10^{-4}$ \\ 
$\sigma_{22}$ [MPa]       & $2.10\cdot10^{-4}$ & $4.34\cdot10^{-4}$ \\ 
$\sigma_{33}$ [MPa]       & $3.35\cdot10^{-4}$ & $6.40\cdot10^{-4}$ \\ 
$\sigma_{12}$ [MPa]       & $6.73\cdot10^{-5}$ & $1.17\cdot10^{-4}$ \\ 
$\sigma_{13}$ [MPa]       & $1.80\cdot10^{-4}$ & $3.24\cdot10^{-4}$ \\ 
$\sigma_{23}$ [MPa]       & $2.95\cdot10^{-5}$ & $5.86\cdot10^{-5}$ \\ \hline
\end{tabular}
\end{table}

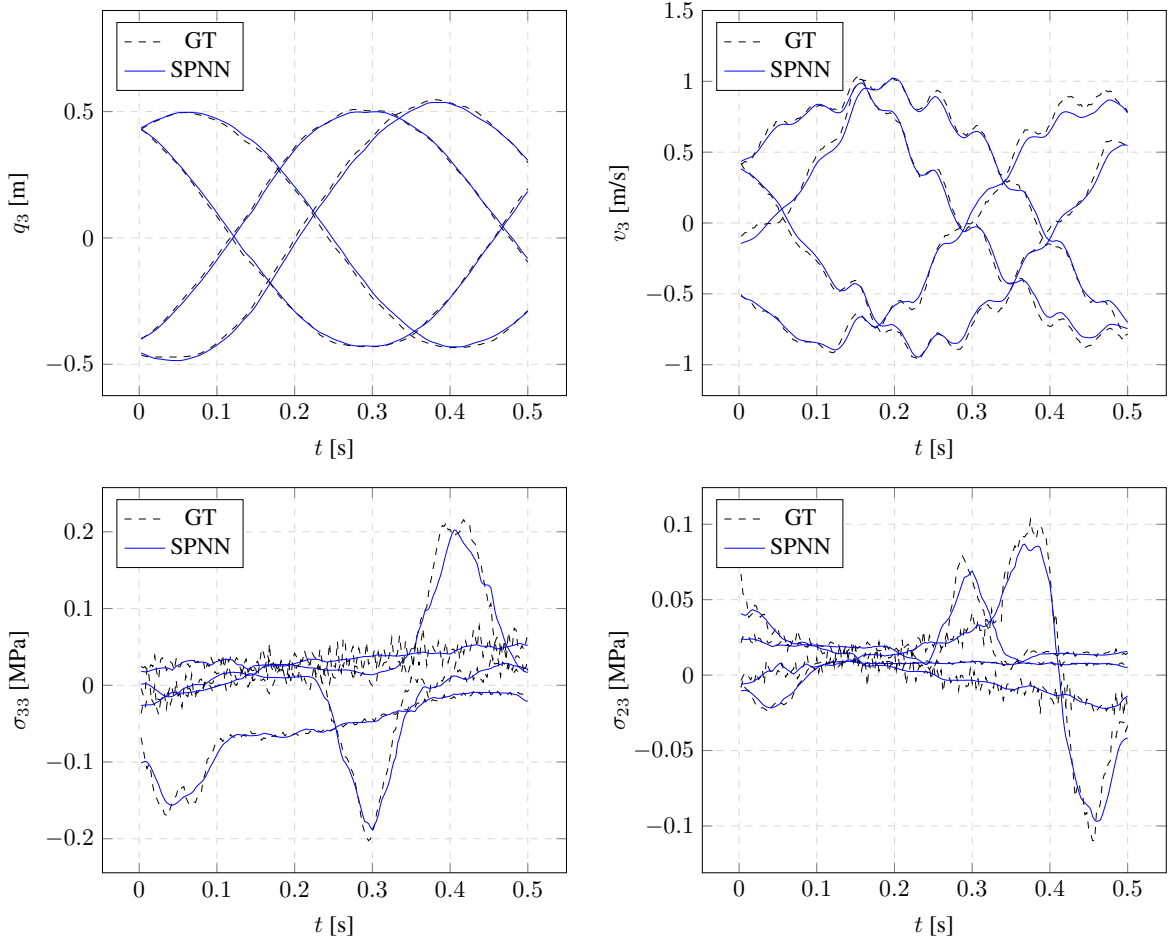
\begin{figure*}[h]
\centering
\begin{tikzpicture}[scale=0.9]

% q
\begin{axis}[name=plot1,
  grid=major, % Display a grid
  grid style={dashed,gray!30}, % Set the style
  xlabel={$t$ [s]},
  ylabel={$q_3$ [m]},
  cycle list name = {color list},
  ymax=0.9, legend pos=north west]
  
  \foreach \F in {1,...,4}{
	\addplot [color=black, dashed] table [y=GT_\F, x=tspan]{graphs/tire/results_q3.txt};
	\ifthenelse{\F=1}{\addlegendentry{GT}}{}
	\addplot [color=blue] table [y=SPNN_\F, x=tspan]{graphs/tire/results_q3.txt};
	\ifthenelse{\F=1}{\addlegendentry{SPNN}}{}
	}
\end{axis}

% v
\begin{axis}[name=plot2, at={($(plot1.right of south east)+(2cm,0)$)},
  grid=major, % Display a grid
  grid style={dashed,gray!30}, % Set the style
  xlabel={$t$ [s]},
  ylabel={$v_3$ [m/s]},
  cycle list name = {color list},
  ymax=1.5, legend pos=north west]
  
  \foreach \F in {1,...,4}{
	\addplot [color=black, dashed] table [y=GT_\F, x=tspan]{graphs/tire/results_v3.txt};
	\ifthenelse{\F=1}{\addlegendentry{GT}}{}
	\addplot [color=blue] table [y=SPNN_\F, x=tspan]{graphs/tire/results_v3.txt};
	\ifthenelse{\F=1}{\addlegendentry{SPNN}}{}
	}	
\end{axis}
	
% sigma 33
\begin{axis}[name=plot3, at={($(plot1.below south west)+(0,-6cm)$)},
  grid=major, % Display a grid
  grid style={dashed,gray!30}, % Set the style
  xlabel={$t$ [s]},
  ylabel={$\sigma_{33}$ [MPa]},
  cycle list name = {color list},
  legend pos=north west]
  
  \foreach \F in {1,...,4}{
	\addplot [color=black, dashed] table [y=GT_\F, x=tspan]{graphs/tire/results_s33.txt};
	\ifthenelse{\F=1}{\addlegendentry{GT}}{}
    \addplot [color=blue] table [y=SPNN_\F, x=tspan]{graphs/tire/results_s33.txt};
	\ifthenelse{\F=1}{\addlegendentry{SPNN}}{}
	}	
\end{axis}

% sigma 23
\begin{axis}[name=plot4, at={($(plot3.right of south east)+(2cm,0)$)},
  grid=major, % Display a grid
  grid style={dashed,gray!30}, % Set the style
  xlabel={$t$ [s]},
  ylabel={$\sigma_{23}$ [MPa]},
  cycle list name = {color list},
  yticklabel style={/pgf/number format/.cd,fixed,precision=2},
  legend pos=north west]
  
  \foreach \F in {1,...,4}{
	\addplot [color=black, dashed] table [y=GT_\F, x=tspan]{graphs/tire/results_s23.txt};
	\ifthenelse{\F=1}{\addlegendentry{GT}}{}
	\addplot [color=blue] table [y=SPNN_\F, x=tspan]{graphs/tire/results_s23.txt};
	\ifthenelse{\F=1}{\addlegendentry{SPNN}}{}
	}	
\end{axis}

\end{tikzpicture}
\caption{Results of the complete integration scheme (SPNN) with respect to the ground truth simulation (GT) for 4 different nodes and 4 different variables ($q_3$, $v_3$, $\sigma_{33}$ and $\sigma_{23}$) of the hyperelastic rolling tire database.}
\label{fig:tire_results}
\end{figure*}

\myfigref{fig:tire_results} presents the time evolution of the decoded state variables $q_3$, $v_3$, $\sigma_{33}$ and $\sigma_{23}$ of the rolling hyperelastic tire for 4 different nodes computed with the presented integration scheme and the ground truth. The mean squared error of the total integration scheme, computed with \myalgref{alg:SPNN_test}, for the 12 state variables is reported in \mytabref{tab:tire_results_SPNN} { using the Structure-preserving neural network (SPNN) and the unconstrained approach (UC). In this example, the error achieved by our method is several orders of magnitude less than the naive approach}.

\begin{table}[h!]

\caption{Mean squared error of the SPNN integration scheme and the unconstrained (UC) approach for the 12 state variables of the rolling tire example, reported for the complete trajectory.}\label{tab:tire_results_SPNN}
\centering
\begin{tabular}{lcc}
\hline
State variable ($\bs{z}_i$) & $\text{MSE}^{\text{SPNN}}$ & $\text{MSE}^{\text{UC}}$ \\ \hline
$q_1$ [m]         	     & $2.07\cdot10^{-4}$ & $2.76\cdot10^{-1}$ \\ 
$q_2$ [m]                 & $8.09\cdot10^{-7}$ & $1.57\cdot10^{-5}$ \\ 
$q_3$ [m]                 & $6.25\cdot10^{-5}$ & $5.75\cdot10^{-2}$ \\ \hline
$v_1$ [m/s]               & $7.95\cdot10^{-3}$ & $4.26$ \\ 
$v_2$ [m/s]               & $3.79\cdot10^{-5}$ & $7.00\cdot10^{-4}$ \\ 
$v_3$ [m/s]               & $1.78\cdot10^{-2}$ & $4.39$ \\ \hline
$\sigma_{11}$ [MPa]       & $2.04\cdot10^{-4}$ & $4.71\cdot10^{-3}$ \\ 
$\sigma_{22}$ [MPa]       & $1.76\cdot10^{-4}$ & $4.07\cdot10^{-3}$ \\ 
$\sigma_{33}$ [MPa]       & $2.70\cdot10^{-4}$ & $5.35\cdot10^{-3}$ \\ 
$\sigma_{12}$ [MPa]       & $6.05\cdot10^{-5}$ & $1.50\cdot10^{-3}$ \\ 
$\sigma_{13}$ [MPa]       & $1.48\cdot10^{-4}$ & $3.03\cdot10^{-3}$ \\ 
$\sigma_{23}$ [MPa]       & $2.73\cdot10^{-5}$ & $6.67\cdot10^{-4}$ \\ \hline
\end{tabular}
\end{table}

\section{Conclusions}\label{sec:conc}

In this work, we { develop a technique to learn the latent dimensionality of a physical system from data and obtain a thermodynamics-aware time integrator, which guarantees the fulfillment of the laws of thermodynamics}. This technique is applied to two different physical systems. The Couette flow in a viscoelastic fluid is reduced from $D=400$ dimensions to $d=4$ dimensions, whereas the rolling tire is reduced from $D=49680$ dimensions to $d=9$ dimensions. The physically informed integrator is then able to predict the full time evolution of the set of state variables with similar precision reported in previous work \cite{hernandez2020structure,gonzalez2018datadriven}.

If compared to previous works of the authors in the field, the use of autoencoders to unveil the dimensionality of the embedding manifold { {clearly}} outperforms the results obtained by classical (linear) model order reduction techniques{, specially in highly nonlinear state variables such as the rolling tire velocity}. In addition, it is worth highlighting the fact that the method is able to detect the true dimensionality of the data, with no need to {{call to different codes} rely on additional methods, such as k-PCA or topological data analysis, for instance} for this purpose. The right thermodynamic setting also ensures the consistency {{and stability}} of the full-order dynamics, after projecting back the reduced-order results to the physical space, {achieving better results than an unconstrained approach with no physical restrictions}. 

Some of the future work, including several improvements to the proposed algorithm, are listed below.
\begin{itemize}
\item \textbf{Database}: A limitation of the present work is the use of synthetic instead of experimental data. A research field is opened to test the limits of the presented methodology applied to real captured data, and to study the influence of noise in the measurements. 

{\item \textbf{Autoencoder latent space}: The latent space of the reduction step can not only be sparsified, but also regularized using variational inference via Variational Autoencoders \cite{kingma2013auto}. This Bayesian approach is convenient in cases where the latent variables are sampled or interpolated, and lead to smoother transitions between them. A future line of this work is to explore VAEs applied to physical systems and study its influence in the latent space topology and extrapolability.}

\item \textbf{Nets Architecture}: The solution of many physical systems has highly spatio-temporal correlations. Thus, convolutional \cite{tompson2016accelerating} and graph-based \cite{zhou2018graph} neural networks could be a more optimized approach, reducing the network complexity and speeding up the learning process. 

{\item \textbf{Energy and Entropy gradients}: A modification can be performed in the structure-preserving neural network in order to output directly the energy and entropy functions. Then, the energy and entropy gradients of the GENERIC formulation can be computed via automatic differentiation with respect to the network input, the state vector. This way, the energy and entropy gradients are forced to be integrable \cite{teichert2019machine}.}

\end{itemize}

\bibliography{ROM-NN-ARXIV}

\begin{thebibliography}{10}

\bibitem{manifold}
Charles Fefferman, Sanjoy Mitter, and Hariharan Narayanan.
\newblock Testing the manifold hypothesis.
\newblock {\em Journal of the American Mathematical Society}, 29(4):983--1049,
  2016.

\bibitem{niroomandi2008real}
Siamak Niroomandi, Ic{\'\i}ar Alfaro, El{\'\i}as Cueto, and Francisco Chinesta.
\newblock Real-time deformable models of non-linear tissues by model reduction
  techniques.
\newblock {\em Computer methods and programs in biomedicine}, 91(3):223--231,
  2008.

\bibitem{du2013pod}
Juan Du, Fangxin Fang, Christopher~C Pain, IM~Navon, Jiang Zhu, and David~A
  Ham.
\newblock Pod reduced-order unstructured mesh modeling applied to 2d and 3d
  fluid flow.
\newblock {\em Computers \& Mathematics with Applications}, 65(3):362--379,
  2013.

\bibitem{prud2002reliable}
Christophe Prud'Homme, Dimitrios~V Rovas, Karen Veroy, Luc Machiels, Yvon
  Maday, Anthony~T Patera, and Gabriel Turinici.
\newblock Reliable real-time solution of parametrized partial differential
  equations: Reduced-basis output bound methods.
\newblock {\em J. Fluids Eng.}, 124(1):70--80, 2002.

\bibitem{rowley2004model}
Clarence~W Rowley, Tim Colonius, and Richard~M Murray.
\newblock Model reduction for compressible flows using pod and galerkin
  projection.
\newblock {\em Physica D: Nonlinear Phenomena}, 189(1-2):115--129, 2004.

\bibitem{farrell2011conservative}
PE~Farrell and JR~Maddison.
\newblock Conservative interpolation between volume meshes by local galerkin
  projection.
\newblock {\em Computer Methods in Applied Mechanics and Engineering},
  200(1-4):89--100, 2011.

\bibitem{badias2019augmented}
Alberto Bad{\'\i}as, Sarah Curtit, David Gonz{\'a}lez, Ic{\'\i}ar Alfaro,
  Francisco Chinesta, and El{\'\i}as Cueto.
\newblock An augmented reality platform for interactive aerodynamic design and
  analysis.
\newblock {\em International Journal for Numerical Methods in Engineering},
  120(1):125--138, 2019.

\bibitem{moya2019learning}
Beatriz Moya, David Gonz{\'a}lez, Ic{\'\i}ar Alfaro, Francisco Chinesta, and
  E~Cueto.
\newblock Learning slosh dynamics by means of data.
\newblock {\em Computational Mechanics}, 64(2):511--523, 2019.

\bibitem{moya2020physically}
Beatriz Moya, Iciar Alfaro, David Gonzalez, Francisco Chinesta, and El{\'\i}as
  Cueto.
\newblock Physically sound, self-learning digital twins for sloshing fluids.
\newblock {\em PLOS ONE}, 15(6):e0234569, 2020.

\bibitem{goodfellow2016deep}
Ian Goodfellow, Yoshua Bengio, and Aaron Courville.
\newblock {\em Deep learning}.
\newblock MIT press, 2016.

\bibitem{farina2020searching}
Marco Farina, Yuichiro Nakai, and David Shih.
\newblock Searching for new physics with deep autoencoders.
\newblock {\em Physical Review D}, 101(7):075021, 2020.

\bibitem{liu2018constrained}
Qi~Liu, Miltiadis Allamanis, Marc Brockschmidt, and Alexander Gaunt.
\newblock Constrained graph variational autoencoders for molecule design.
\newblock In {\em Advances in neural information processing systems}, pages
  7795--7804, 2018.

\bibitem{lee2020model}
Kookjin Lee and Kevin~T Carlberg.
\newblock Model reduction of dynamical systems on nonlinear manifolds using
  deep convolutional autoencoders.
\newblock {\em Journal of Computational Physics}, 404:108973, 2020.

\bibitem{marco2017deeptof}
Julio Marco, Quercus Hernandez, Adolfo Munoz, Yue Dong, Adrian Jarabo, Min~H
  Kim, Xin Tong, and Diego Gutierrez.
\newblock Deeptof: off-the-shelf real-time correction of multipath interference
  in time-of-flight imaging.
\newblock {\em ACM Transactions on Graphics (ToG)}, 36(6):1--12, 2017.

\bibitem{raissi2019physics}
Maziar Raissi, Paris Perdikaris, and George~E Karniadakis.
\newblock Physics-informed neural networks: A deep learning framework for
  solving forward and inverse problems involving nonlinear partial differential
  equations.
\newblock {\em Journal of Computational Physics}, 378:686--707, 2019.

\bibitem{kelly2020easynode}
Jacob Kelly, Jesse Bettencourt, Matthew~James Johnson, and David Duvenaud.
\newblock Learning differential equations that are easy to solve.
\newblock In {\em Neural Information Processing Systems}, 2020.

\bibitem{bertalan2019learning}
Tom Bertalan, Felix Dietrich, Igor Mezi{\'c}, and Ioannis~G Kevrekidis.
\newblock On learning hamiltonian systems from data.
\newblock {\em Chaos: An Interdisciplinary Journal of Nonlinear Science},
  29(12):121107, 2019.

\bibitem{greydanus2019hamiltonian}
Samuel Greydanus, Misko Dzamba, and Jason Yosinski.
\newblock Hamiltonian neural networks.
\newblock In {\em Advances in Neural Information Processing Systems}, pages
  15379--15389, 2019.

\bibitem{toth2019hamiltonian}
Peter Toth, Danilo~Jimenez Rezende, Andrew Jaegle, S{\'e}bastien Racani{\`e}re,
  Aleksandar Botev, and Irina Higgins.
\newblock Hamiltonian generative networks.
\newblock {\em arXiv preprint arXiv:1909.13789}, 2019.

\bibitem{zhong2019symplectic}
Yaofeng~Desmond Zhong, Biswadip Dey, and Amit Chakraborty.
\newblock Symplectic ode-net: Learning hamiltonian dynamics with control.
\newblock {\em arXiv preprint arXiv:1909.12077}, 2019.

\bibitem{tong2020symplectic}
Yunjin Tong, Shiying Xiong, Xingzhe He, Guanghan Pan, and Bo~Zhu.
\newblock Symplectic neural networks in taylor series form for hamiltonian
  systems.
\newblock {\em arXiv preprint arXiv:2005.04986}, 2020.

\bibitem{jin2020learning}
Pengzhan Jin, Zhen Zhang, Ioannis~G Kevrekidis, and George~Em Karniadakis.
\newblock Learning poisson systems and trajectories of autonomous systems via
  poisson neural networks.
\newblock {\em arXiv preprint arXiv:2012.03133}, 2020.

\bibitem{ottinger1997dynamics}
Hans~Christian {\"O}ttinger and Miroslav Grmela.
\newblock Dynamics and thermodynamics of complex fluids. ii. illustrations of a
  general formalism.
\newblock {\em Physical Review E}, 56(6):6633, 1997.

\bibitem{grmela1997dynamics}
Miroslav Grmela and Hans~Christian {\"O}ttinger.
\newblock Dynamics and thermodynamics of complex fluids. i. development of a
  general formalism.
\newblock {\em Physical Review E}, 56(6):6620, 1997.

\bibitem{hernandez2020structure}
Quercus Hernandez, Alberto Badias, David Gonzalez, Francisco Chinesta, and
  Elias Cueto.
\newblock Structure-preserving neural networks.
\newblock {\em arXiv preprint arXiv:2004.04653}, 2020.

\bibitem{E2017}
Weinan E.
\newblock A proposal on machine learning via dynamical systems.
\newblock {\em Communications in Mathematics and Statistics}, 5(1):1--11, Mar
  2017.

\bibitem{ottinger2015preservation}
Hans~Christian {\"O}ttinger.
\newblock Preservation of thermodynamic structure in model reduction.
\newblock {\em Physical Review E}, 91(3):032147, 2015.

\bibitem{ng2011sparse}
Andrew Ng et~al.
\newblock Sparse autoencoder.
\newblock {\em CS294A Lecture notes}, 72(2011):1--19, 2011.

\bibitem{paszke2017automatic}
Adam Paszke, Sam Gross, Soumith Chintala, Gregory Chanan, Edward Yang, Zachary
  DeVito, Zeming Lin, Alban Desmaison, Luca Antiga, and Adam Lerer.
\newblock Automatic differentiation in pytorch.
\newblock {\em Autodiff Workshop: The Future of Gradient-based Machine Learning
  Software and Techniques}, 2017.

\bibitem{ruder2016overview}
Sebastian Ruder.
\newblock An overview of gradient descent optimization algorithms.
\newblock {\em arXiv preprint arXiv:1609.04747}, 2016.

\bibitem{battaglia2018relational}
Peter~W Battaglia, Jessica~B Hamrick, Victor Bapst, Alvaro Sanchez-Gonzalez,
  Vinicius Zambaldi, Mateusz Malinowski, Andrea Tacchetti, David Raposo, Adam
  Santoro, Ryan Faulkner, et~al.
\newblock Relational inductive biases, deep learning, and graph networks.
\newblock {\em arXiv preprint arXiv:1806.01261}, 2018.

\bibitem{morrison1986paradigm}
Philip~J Morrison.
\newblock A paradigm for joined hamiltonian and dissipative systems.
\newblock {\em Physica D: Nonlinear Phenomena}, 18(1-3):410--419, 1986.

\bibitem{schmidhuber2015deep}
J{\"u}rgen Schmidhuber.
\newblock Deep learning in neural networks: An overview.
\newblock {\em Neural networks}, 61:85--117, 2015.

\bibitem{laso1993calculation}
Manuel Laso and Hans~Christian {\"O}ttinger.
\newblock Calculation of viscoelastic flow using molecular models: the
  connffessit approach.
\newblock {\em Journal of Non-Newtonian Fluid Mechanics}, 47:1--20, 1993.

\bibitem{le2009multiscale}
Claude Le~Bris and Tony Lelievre.
\newblock Multiscale modelling of complex fluids: a mathematical initiation.
\newblock In {\em Multiscale modeling and simulation in science}, pages
  49--137. Springer, 2009.

\bibitem{he2015delving}
Kaiming He, Xiangyu Zhang, Shaoqing Ren, and Jian Sun.
\newblock Delving deep into rectifiers: Surpassing human-level performance on
  imagenet classification.
\newblock In {\em Proceedings of the IEEE international conference on computer
  vision}, pages 1026--1034. ICCV, 2015.

\bibitem{kingma2014adam}
Diederik~P Kingma and Jimmy Ba.
\newblock Adam: A method for stochastic optimization.
\newblock {\em arXiv preprint arXiv:1412.6980}, 2014.

\bibitem{gonzalez2019thermodynamically}
David Gonz{\'a}lez, Francisco Chinesta, and El{\'\i}as Cueto.
\newblock Thermodynamically consistent data-driven computational mechanics.
\newblock {\em Continuum Mechanics and Thermodynamics}, 31(1):239--253, 2019.

\bibitem{gonzalez2018datadriven}
D.~Gonz{\'a}lez, F.~Chinesta, and E.~Cueto.
\newblock Consistent data-driven computational mechanics.
\newblock {\em AIP Conference Proceedings}, 1960(1):090005, 2018.

\bibitem{kingma2013auto}
Diederik~P Kingma and Max Welling.
\newblock Auto-encoding variational bayes.
\newblock {\em arXiv preprint arXiv:1312.6114}, 2013.

\bibitem{tompson2016accelerating}
Jonathan Tompson, Kristofer Schlachter, Pablo Sprechmann, and Ken Perlin.
\newblock Accelerating eulerian fluid simulation with convolutional networks.
\newblock {\em arXiv preprint arXiv:1607.03597}, 2016.

\bibitem{zhou2018graph}
Jie Zhou, Ganqu Cui, Zhengyan Zhang, Cheng Yang, Zhiyuan Liu, Lifeng Wang,
  Changcheng Li, and Maosong Sun.
\newblock Graph neural networks: A review of methods and applications.
\newblock {\em arXiv preprint arXiv:1812.08434}, 2018.

\bibitem{teichert2019machine}
Gregory~H Teichert, AR~Natarajan, A~Van~der Ven, and Krishna Garikipati.
\newblock Machine learning materials physics: Integrable deep neural networks
  enable scale bridging by learning free energy functions.
\newblock {\em Computer Methods in Applied Mechanics and Engineering},
  353:201--216, 2019.

\end{thebibliography}
\bibliographystyle{unsrt}

\end{document}